\documentclass{aa}

\usepackage{xspace}
\newcommand{\athenapk}{{\scshape AthenaPK}\xspace}
\newcommand{\parthenon}{{\scshape Parthenon}\xspace}
\newcommand{\xmagnet}{{\scshape XMagnet}\xspace}

\usepackage{comment}
\usepackage{natbib}
\bibpunct{(}{)}{;}{a}{}{,}

\usepackage[colorlinks=true, urlcolor=blue, colorlinks=true, citecolor=blue, linkcolor=blue]{hyperref}
\usepackage{graphicx}
\usepackage{txfonts}

\newcommand{\red}[1]{\textcolor{red}{#1}}
\renewcommand{\red}[1]{#1}

\begin{document} 

   \title{XMAGNET -- Stir before serving: a Lagrangian perspective on mixing-driven condensation in the intracluster medium}
   
\author{M. Fournier\inst{1}, P. Grete\inst{1}, M. Brüggen\inst{1},  
           B. W. O'Shea\inst{2,3,4,5}, \\
           G. M. Voit\inst{3},
           B. D. Wibking\inst{3},
           D. Prasad\inst{6}
          }

\institute{Universität Hamburg, Hamburger Sternwarte, Gojenbergsweg 112, 21029 Hamburg, Germany \\
              \email{martin.fournier@uni-hamburg.de}
        \and
    Department of Computational Mathematics, Science, and Engineering, Michigan State University, East Lansing, MI 48824, USA
         \and
    Department of Physics and Astronomy, Michigan State University, East Lansing, MI 48824, USA
        \and
    Facility for Rare Isotope Beams, Michigan State University, East Lansing, MI 48824, USA
        \and
    Institute for Cyber-Enabled Research, 567 Wilson Road, Michigan State University, East Lansing, MI 48824
        \and
    School of Physics and Astronomy, Cardiff University, 5 The Parade, Cardiff CF24 3AA, UK
}

   \date{Received \today}
     
    \abstract
    {The condensation of gas in the intracluster medium (ICM) remains a key problem in understanding the thermal regulation of galaxy cluster cores. Magnetic fields are expected to play an important role, but their impact is not well understood.}
    {We aim to characterize the thermodynamic and dynamical conditions leading to condensation in cluster cores, and to assess the role of magnetic fields.}
    {We implement a Monte-Carlo tracer particle algorithm in the GPU-accelerated code \textsc{AthenaPK}, and run a purely hydrodynamical and a magnetohydrodynamical (MHD) simulations of an idealized cool-core cluster. We identify the subset of hot ICM tracers that undergo a transition to the cold phase and reconstruct their histories over a lookback time of $300\,\mathrm{Myr}$ prior to condensation.}
    {In both runs, the large majority of tracers transitioning to the cold phase follow a thermodynamic pathway driven by mixing, whereby hot ambient gas is entrained onto low-entropy seed clumps that subsequently grow into larger clouds and filaments. In the hydrodynamical run, these seeds form mainly via in-situ cooling at the edges of AGN cavities. In the MHD run, the cold gas cycle is more complex: AGN outflows occasionally shred portions of existing filaments into fragments which are then uplifted, seeding further condensation. In the MHD run, the properties of condensing tracers begin to diverge from the background ICM significantly earlier than in the hydrodynamical run (${\sim}150\,\rm Myr$ before the cooling transition versus ${\sim}30\,\rm Myr$), with vorticity and magnetic energy growing together. The turbulent Mach number at condensation is also systematically lower than in the hydrodynamical run. We examine the post-condensation evolution of individual cold structures in the MHD run, namely a massive core filament and two isolated clouds in quiescent regions. We find that magnetic tension dominates over ram pressure as the primary drag force, significantly reducing the clouds' terminal velocity.}
    {Our results demonstrate that magnetic fields substantially impact the assembly history and kinematic properties of the cold phase in cool--core clusters.}

   \titlerunning{Mixing-Driven Condensation in the ICM}
   \authorrunning{M. Fournier et al.} 
   \maketitle
   
\setlength{\parindent}{0cm}

\section{Introduction}

Observations across a wide range of astrophysical systems indicate that gaseous halos often exist in multiple thermodynamic phases, spanning orders of magnitude in density and temperature. In particular, clouds and filaments of atomic and molecular gas have been found to pervade the circumgalactic medium (CGM) of star forming galaxies \citep[e.g.,][]{Lynds_1963,Lanzetta_1995,Chen_1998,Faucher_2023}, the innermost regions of cool-core galaxy cluster's intracluster medium (ICM) \citep[e.g.,][]{Lynds_1970,Salome_2006,Olivares_2019,Donahue_2022}, and the halo of distant quasars \citep[e.g.,][]{Arrigoni_2019}. This cold ($T \sim 10-10^4$ K) component is thought to play a key role in the thermal regulation of these systems, as it provides a large reservoir of material for star formation, supermassive black hole (SMBH) feeding and subsequently active galactic nuclei (AGN) feedback \citep[e.g.][]{Gaspari_2013,Guo_2024}. A key question is under what conditions this cold phase condenses out of the hot ambient medium. This is commonly framed in terms of the competition between radiative cooling and gravitational potential, quantified by the cooling time $t_{\rm{cool}}$ and free-fall time $t_{\rm{ff}}$, defined as:
\begin{equation}
t_{\rm{cool}} = \frac{3 n k_{\rm{B}} T}{2 n_{\rm{H}}^2 \Lambda(T)}, \qquad
t_{\rm{ff}} = \sqrt{\frac{2r}{g(r)}},
\end{equation}
where $n$ is the gas number density, $n_{\rm{H}}$ is the hydrogen number density, $T$ is the temperature, $\Lambda(T)$\footnote{The cooling function $\Lambda(T)$ used in this work is tabulated per $n_{\rm{H}}^2$, as opposed to the more common $n_e n_{\rm{H}}$ convention.} is the cooling function, $r$ is the radial distance from the center, and $g(r)$ is the gravitational acceleration at radius $r$. Deprojected X-ray observations of nearby galaxy clusters indicate that cores containing more than $10^8 \, \rm{M}_\odot$ of cold gas are characterized by $t_{\rm{cool}} \lesssim 1$ Gyr and $10 \lesssim \min(t_{\rm{cool}}/t_{\rm{ff}}) \lesssim 30$, see e.g., \citet{Hogan_2017, Babyk_2018} and Sect. 8.1 of \citet{Donahue_2022} for a recent review. 

How multiphase gas forms at values well above $t_{\rm{cool}}/t_{\rm{ff}} \sim 1$ remains an open question. Turbulence alone seems insufficient to generate precipitation in idealized setups for $t_{\rm{cool}}/t_{\rm{ff}} \gtrsim 5$ \citep[e.g.][]{McCourt_2012, Gaspari_2013}, even with external turbulent driving \citep{Wibking_2024}. One pathway that has received considerable attention is AGN-driven uplift: theoretical work and hydrodynamical simulations of self-regulated AGN feedback in radially stratified atmospheres have shown that jets can displace low-entropy gas to sufficiently large radii that $t_{\rm{cool}}/t_{\rm{ff}}$ approaches unity, promoting its condensation into cold clumps and filaments \citep{Revaz_2008,Li_clumps,McNamara_2016,Voit_2017,Hillel_2018,Choudhury_2019,Zhang_2022}.

Recent work \citep[see e.g.][]{Ji_2018,Wibking_2025,Voit_2026} shown that magnetic fields strongly promote precipitation, with magnetized plasmas being much more prone to condensation than unmagnetized ones. Their results show that modest field strengths ($\beta \sim 10^2$) substantially raise the critical $t_{\rm{cool}}/t_{\rm{ff}}$ threshold for the onset of multiphase gas, reaching values around $t_{\rm{cool}}/t_{\rm{ff}} \sim 10$ even in the absence of any externally driven turbulence. A possible interpretation of this enhancement is the effect of magnetic tension acceleration $\mathbf{a}_{\rm B}^{\tau} \propto (\mathbf{B} \cdot \nabla)\mathbf{B}/\rho$. In an unmagnetized atmosphere, thermally unstable perturbations couple to buoyant oscillations at the Brunt-V\"{a}is\"{a}l\"{a} frequency, saturating before reaching nonlinear density contrasts when $t_{\rm cool}/t_{\rm ff} \gtrsim 1$ \citep{McCourt_2012}. Weak magnetic fields introduce magnetothermal drips \citep{Voit_2026}, where magnetic tension nearly offsets buoyancy, allowing low-entropy gas to descend and cool on a timescale $\sim t_{\rm cool}$. As low-entropy gas falls, flux freezing stretches field lines into a kink whose tension slows the descent to sub-Keplerian speeds while amplifying the local field \citep{Voit_2026}. Numerous previous studies have reported individual signatures of this effect, including cloud-crushing \citep{Kaul_2025, Cottle2020}, stratified box \citep{Ji_2018,Wibking_2025}, and idealized cluster core \citep{Wang_2020,Wang_2021,Ehlert_2023,Fournier_2024b} simulations. In \citet{Wang_2021}, the overall magnetic tension averaged over the whole bulk mass of a forming filament is found to exceed all other acceleration terms a few Myr after its formation. 

These magnetic effects, however, may be modulated by the complexity of the ambient environment. In the context of cold clouds in the CGM, \citet{ramesh2026} (see also \citet{Ramesh_MC}) find that radiative cooling dominates cloud survival over magnetic effects, with cloud kinematics shaped by a combination of drag deceleration and momentum exchange with the background velocity field. In contrast, \citet{Hidalgo_2024} find that magnetic fields significantly enhance cloud survival in galactic winds, highlighting the sensitivity of magnetic effects to the local environment. Beyond magnetic fields, turbulence \citep{Ghosh_2025}, compression \citep{Sotira_2026}, anisotropic conduction and cosmic rays \citep{Sharma_2010,Beckmann_2022,Jennings_2023} and cosmological context \citep{Rohr_2025,Staffehl_2025} have been found to play an important role in the formation and distribution of multiphase gas, while the effect of viscosity appears more limited \citep{Marin_2025}.

In this work, we carry out high-resolution MHD simulations of a Perseus-like cluster core with self-regulated AGN feedback and radiative cooling, employing Lagrangian tracer particles to disentangle the contribution of the various condensation channels. This approach allows us to follow gas parcels without any a priori assumption about the conditions that will later lead to condensation. Sec.~\ref{sect:method} describes the simulation setup, followed by the presentation of our results in Sec.~\ref{sect:results}, their interpretation and discussion in Sec.~\ref{sect:discussion}, and our conclusions in Sec.~\ref{sect:conclusion}.

\begin{figure*}[h!]
\includegraphics[width=\textwidth]{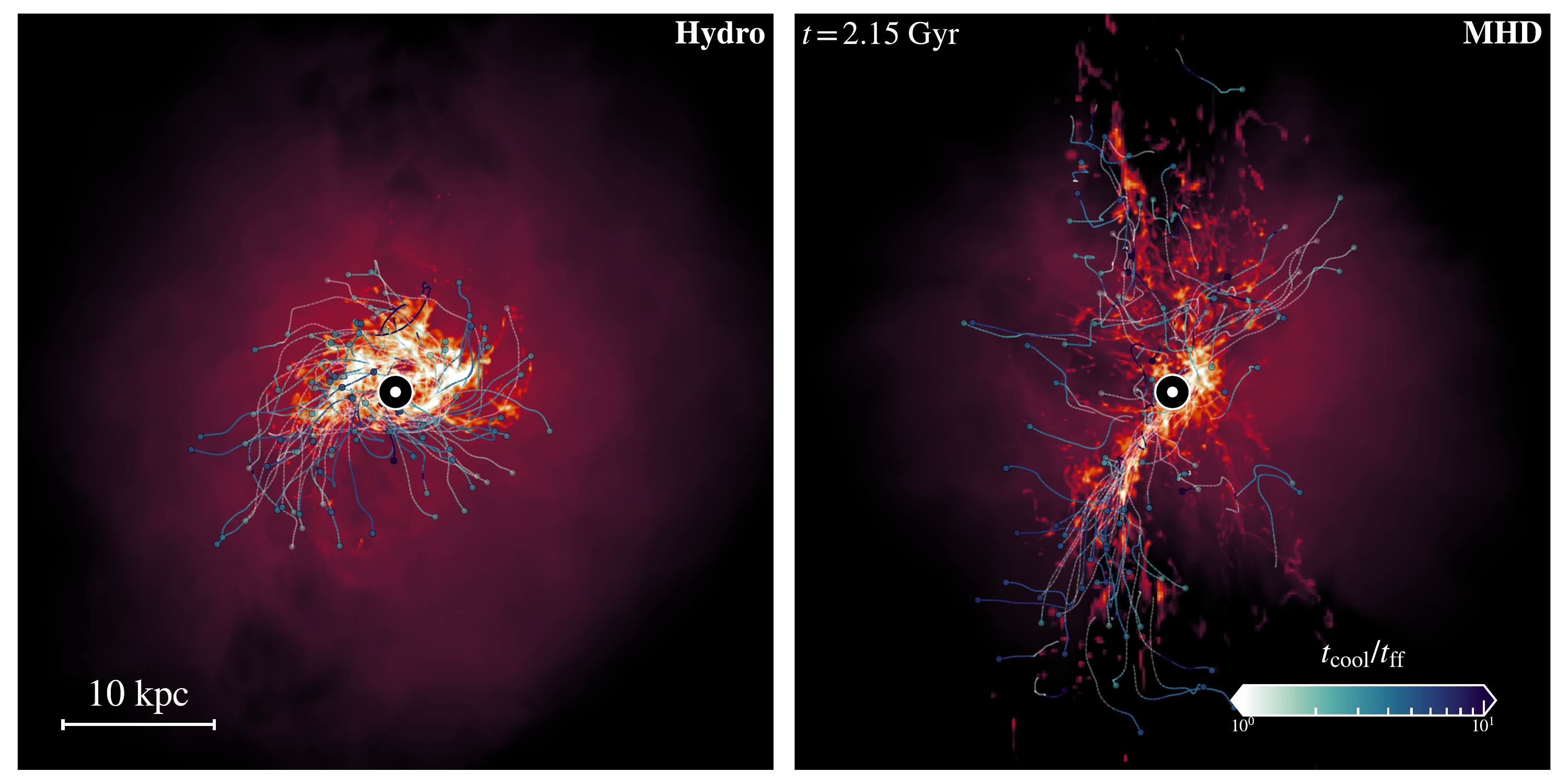}
    \caption{Density projections of the innermost 50 kpc of the purely hydrodynamic (left) and MHD (right) simulations at $t = 2.15$ Gyr. Overlaid are the trajectories of 100 randomly selected tracer particles undergoing a cooling transition over the 50 Myr preceding the snapshot used for the projection, with circular markers indicating each tracer's starting position. Trajectory segments are colored by the local ratio of the cooling time to the free-fall time. The black circular marker indicates the position of the SMBH, the jet axis pointing vertically (see also associated \href{https://youtu.be/eIKyCpKgpsY}{movie}).}
    \label{fig:mainfig}
\end{figure*}

\section{Method}
\label{sect:method}

The simulation setup generally follows the \xmagnet simulation suite \footnote{See \href{https://xmagnet-simulations.github.io/}{https://xmagnet-simulations.github.io/} for further material.} introduced in \citet{Grete_2025}.
We ran hydro and MHD simulations of an idealized cool-core cluster using {\textsc{\athenapk}}\footnote{\textsc{\athenapk} is available at \url{https://github.com/parthenon-hpc-lab/athenapk} and commit \texttt{f589221} was used for the simulations; input files are provided as supplemental material.}, a performance-portable version of the {\scshape{Athena++}} astrophysical magnetohydrodynamics code \citep{Stone_2020} based on the block-structured refinement framework \parthenon \citep{Parthenon}. 
The simulations used an overall second-order accurate, shock-capturing, finite-volume scheme consisting of 
second-order Van Leer time integration, piecewise-linear reconstruction, and an HLLC (HLLD) Riemann solver \citep{Miyoshi_2005} for the pure hydro (MHD) run, respectively. 

Gravity is treated as an acceleration term parametrized by a superposition of a Navarro-Frenk-White (NFW) profile \citep{Navarro_1997}, a Hernquist potential representing the brightest central galaxy's mass \citep{Hernquist_1990}, and a central point-mass representing  the supermassive black hole (SMBH) in the Perseus cluster's BCG. The initial gas distribution is calculated to obey hydrostatic equilibrium assuming the entropy profile of the Perseus cluster from the ACCEPT catalogue \citep{ACCEPT_2009}. 
Optically thin cooling is calculated by the exact integration method introduced by \citet{Townsend_2009} and uses solar metallicity cooling tables by \cite{Schure} down to $10^{4.2}$ K. A temperature floor of $10^{4}$ K was also enforced. Purely solenoidal velocity and magnetic perturbations are seeded at $t=0$ to break the symmetry of the system, with respective dispersion of $\delta v = 75 \, \mathrm{km}\,\mathrm{s}^{-1}$ and $\delta B=1 \, \mu\mathrm{G}$. These perturbations are generated using an inverse parabolic spectrum with a peak scale of 100 kpc. AGN feedback is modeled via source terms for thermal, kinetic and magnetic energy. Cold material (defined as gas cells of temperature $T \leq 5 \times 10^4\,\rm K$ within an accretion radius $r_{\rm acc} = 1\,\rm kpc$) is accreted at a rate $\dot{M}_{\mathrm{acc}}$ related to the accretion power by $\dot{E}_{\mathrm{acc}} = \eta \dot{M}_{\mathrm{acc}} c^2$, where $\eta=0.001$ is the accretion efficiency and $c$ the speed of light. The accretion power is divided among three feedback channels, with 74\% in kinetic energy (jets), 1\% in magnetic fields carried by the jets, and 25\% in thermal feedback. In the absence of star particles, stellar feedback is calculated analytically, by extracting gas denser than a density threshold $n_{\mathrm{thresh.,\star}} = 50 \, \mathrm{cm}^{-3}$ and redistributing it into thermal energy with an efficiency of $5\times10^{-6}$. A detailed description of the full \xmagnet model and parameters is given in \citet{Grete_2025}. The structure of the grid in our run is kept constant across time. The root mesh grid is made of $256^3$ cells covering a total volume of $(3.2$\,Mpc$)^3$. The root grid is refined a further $\ell_{\mathrm{max}} = 7$ cubic and nested refinement levels. As a result, the maximally refined region has a volume of $(25 \mathrm{\, kpc})^3$ and is covered by $256^3$ cells of size $\Delta x = 97.7$ pc.

Each simulation was run for 2.6 Gyr, allowing the first 2 Gyr for the relevant quantities (e.g. radial profiles of entropy and magnetic fields strength) to reach a steady state, after which $10^8$ Lagrangian tracer particles were injected and followed for the remainder of the simulation. We store $\sim 600$ snapshots separated by 1 Myr allowing a detailed analysis at high temporal resolution. The tracer particles are advected following a Monte-Carlo scheme \citep[see e.g.][]{Genel_2013,Cadiou_2019}, as described in Appendix~\ref{app:tracersmethod}.

\section{Results}
\label{sect:results}

\subsection{Properties of the cooling tracers}
\label{sect:globaltracers}

In this subsection, we study the properties of tracer particles undergoing a transition between the hot ($T > 5\times10^6\,\rm K$) and the cold ($T < 10^5\,\rm K$) phase. As a visual introduction, Fig.~\ref{fig:mainfig} shows density projections of the innermost $50\,\rm kpc$ of both runs, overlaid with the trajectories of $100$ randomly selected tracers that underwent a cooling transition within the previous $50\,\rm Myr$, colored by $t_{\rm cool}/t_{\rm ff}$. The associated \href{https://youtu.be/eIKyCpKgpsY}{movie}\footnote{Best viewed at 1080p or higher; quality can be adjusted via the settings menu.} provides a more informative view of the dynamics. \red{In the purely hydrodynamical run, cold gas predominantly settles into a central disk of radius $r \sim 5\,\rm kpc$, fed by structureless clumps forming in-situ at the edges of outflowing AGN cavities; no filamentary structures are present, and the surrounding cold gas forms a disordered distribution with no clear large-scale organization}. These clumps do not appear to grow significantly, as they are efficiently shredded and shattered, possibly through their differential motion with respect to the background ICM or through interactions with the AGN jets. In the MHD run, the picture is more complex: AGN outflows sporadically interact with infalling cold filaments, shredding and uplifting portions of their structure. The resulting fragments subsequently grow and coagulate, forming successive generations of filaments. Additionally, traces of in-situ cooling are visible in the equatorial plane, in regions seemingly undisturbed by AGN jets where the gas may cool more steadily, consistent with similar structures found in \citet{Guo_2024}. These regions produce a rain of infalling clumps toward the accretion zone, potentially sustaining the reservoir of cold gas seeds that are uplifted by AGN jets and later grow through mixing-driven condensation. Overall, gas in the MHD run does not form a kpc-scale disk, instead falling radially toward the center, consistent with recent observations of a $\sim$100~pc circumnuclear disk in NGC~1275 fed by infalling filaments \citep{Oosterloo_2023}, and with theoretical work showing that magnetic fields suppress rotational structures to scales of a few tens of pc \citep{Guo_2024}.

\subsubsection{Cooling mechanism}
\label{sect:coolingmechanism}

Multiple channels can lead to cold gas formation and growth in the ICM and CGM, including direct radiative cooling, mixing-driven cooling, and adiabatic expansion. In this subsection, we estimate which of these channels dominates in our simulations. Adiabatic expansion can be ruled out immediately, as it requires a decrease in density, whereas the cold phase we study consists of overdense clumps formed from hot ICM gas undergoing substantial compression. We therefore focus on distinguishing between direct radiative condensation and mixing-driven cooling.

Gas cooling purely via radiative processes is expected to follow a predictable entropy evolution given by

\begin{equation}
\label{eq:isobaricentropy}
\frac{\mathrm{d}\ln K}{\mathrm{d}t} = -\frac{1}{t_{\rm cool}(K)},
\end{equation}
where $K \equiv k_B T n_e^{-2/3}$ is the ICM entropy, $T$ is the temperature, and $n_e$ is the electron number density. To predict the entropy trajectory of purely radiatively cooling gas, we assume an initially cold state characterized by temperature $T_0$ and pressure $P_0$, and integrate Eq.~\ref{eq:isobaricentropy} backward in time assuming isobaric evolution.

\begin{figure}[h!]
\includegraphics[width=0.5\textwidth]{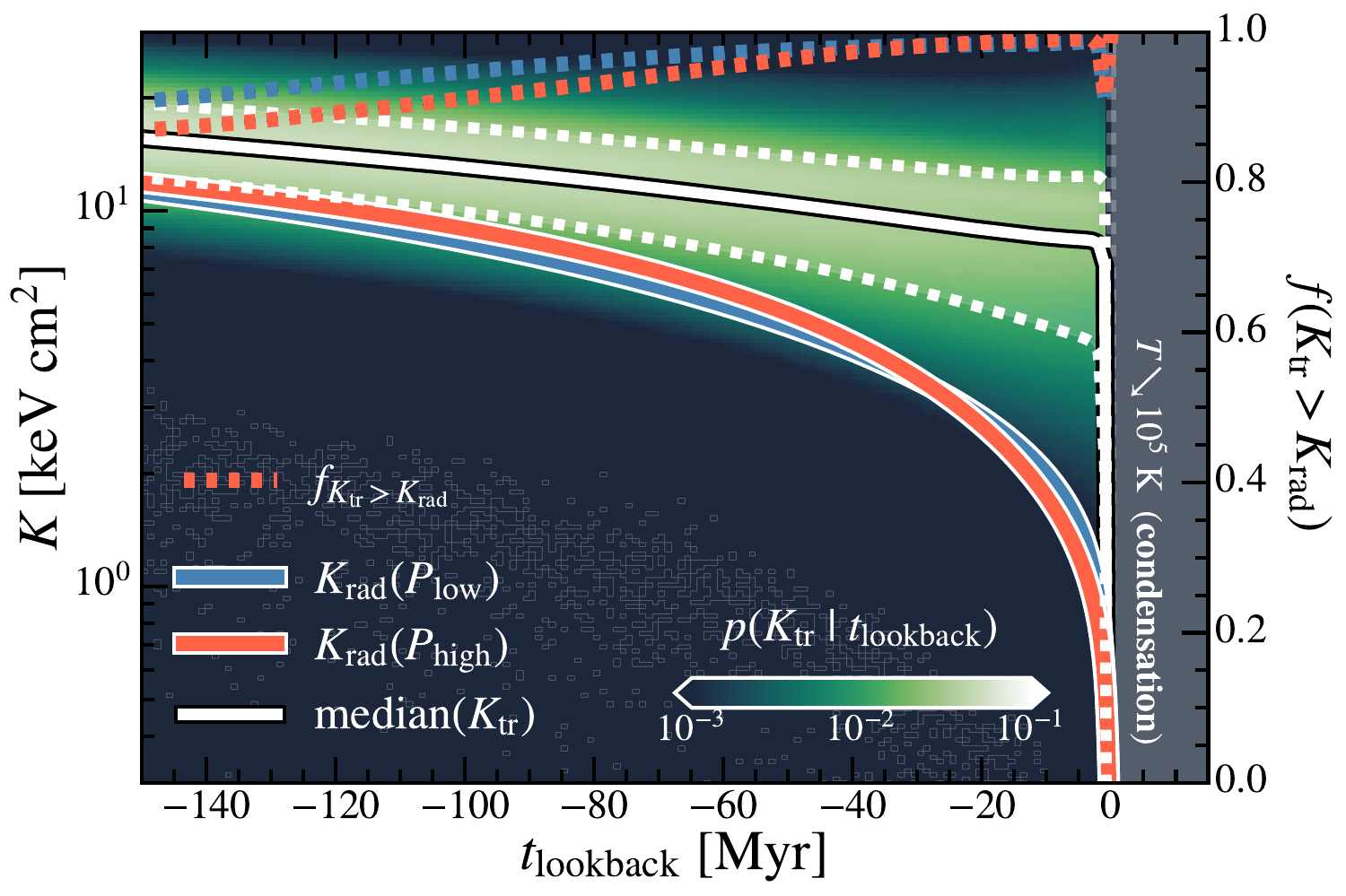}
    \caption{Entropy evolution of tracer particles prior to condensation for our pure hydro run. The 2D histogram shows the distribution of individual tracer entropies $K_{\rm tr}$ as a function of lookback time, with cooling events aligned at $t_{\rm lookback}=0$. The white solid and dashed lines indicate the median and 16th–84th percentile range. Blue and red curves show the isobaric radiative cooling tracks $K_{\rm rad}$ at $P_{\rm low}=4\times10^{-11}$ and $P_{\rm high}=10^{-10}\,\mathrm{dyne\,cm^{-2}}$. Dashed lines (right axis) give the fraction of tracers with $K_{\rm tr} > K_{\rm rad}$.}
    \label{fig:entropy_fraction}
\end{figure}

In parallel, we identify all tracer particles that are initially hot ($T \geq 5 \times 10^6\,\rm K$) and subsequently transition to the cold phase ($T \leq 10^5\,\rm K$) in our simulations. From this set, we randomly select $10^6$ of these tracers (roughly $10\%$ of the full sample), and align their entropy trajectories such that their condensation time, defined as the first snapshot where $T < 5\times10^5$ K, coincides, allowing us to analyze their statistical lookback history.

Fig.~\ref{fig:entropy_fraction} shows a two-dimensional histogram of the entropy evolution $K_{\rm tr}$ of these tracers, as a function of lookback time before condensation, $t_{\rm lookback}$, in the pure hydrodynamic run. The corresponding distribution for the MHD run is nearly indistinguishable, and all conclusions apply equally to both cases. The median evolution is indicated by the solid white line, with 16th-84th percentile range shown as dashed white lines. We estimate the pressure range of the cold phase from the 16th and 84th percentiles of its distribution, yielding $P_{\rm low} = 4 \times 10^{-11}\,\mathrm{dyne\,cm^{-2}}$ and $P_{\rm high} = 10^{-10}\,\mathrm{dyne\,cm^{-2}}$, respectively. Using these values, we integrate Eq.~\ref{eq:isobaricentropy} backward in time with a Runge–Kutta–Fehlberg solver to obtain $K_{\rm rad}$. The resulting curves, shown as the blue and red solid lines, represent the expected entropy evolution of gas parcels cooling purely via radiative processes. Finally, the colored dashed lines indicate, at each lookback time, the fraction of tracers with entropy exceeding the corresponding pure radiative cooling trajectories. More than 80\% of the tracers lie above the expected radiative cooling curve at all times. In addition, at low lookback times ($\vert t_{\rm lookback}\vert = 1$--$2$ Myr), the tracer entropy trajectories typically exhibit a sharp drop from $K \sim 10\,\rm{keV}\,\rm{cm}^2$ down to $K \sim 10^{-2}\,\rm{keV}\,\rm{cm}^2$. Gas located above the radiative trajectory has higher entropy than expected for purely radiative cooling at the same pressure, implying that it has experienced additional heating or mixing with higher-entropy ambient gas. The fact that the large majority of tracers follow this behavior indicates that most gas parcels do not cool through direct radiative condensation, but instead undergo mixing-driven cooling.

\begin{figure}[h!]
\includegraphics[width=0.5\textwidth]{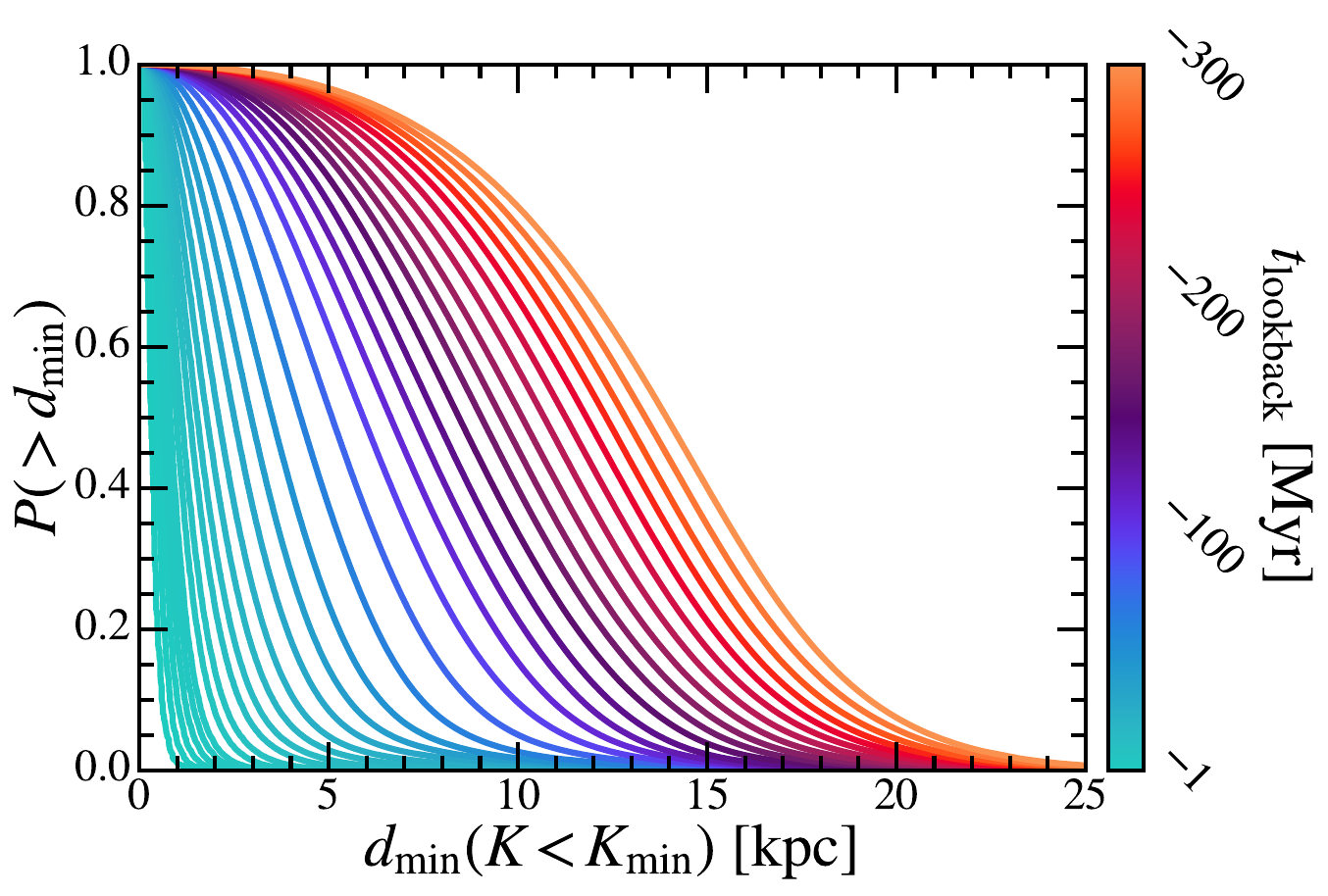}
    \caption{Cumulative distribution of distances $d_{\rm min}$ to the nearest low-entropy cell (defined by $K \leq K_{\rm min} = 1\,\rm{keV}\,\rm{cm}^{2}$) for condensing tracers, evaluated at different lookback times prior to cooling in our pure hydro run. Each curve shows the fraction of tracers with distances greater than $d_{\rm min}$, with color indicating the lookback time.}
    \label{fig:closest_neighbor}
\end{figure}

In Fig.~\ref{fig:closest_neighbor}, we show the evolution of the distance between tracers, at lookback time $t_{\rm lookback}$ prior to condensation, and the nearest cell with entropy below $K_{\rm min} = 1\,\mathrm{keV}\,\mathrm{cm}^{2}$, used as a proxy for the pre-existing cold phase. This distance converges to the local spatial resolution limit $\Delta x$ as cooling transition time approaches, indicating that a significant fraction of the condensing hot gas approaches pre-existing cold structures in the lead-up to cooling. We have verified that our results are robust to the exact choice of $K_{\rm min}$, and that the same conclusions hold for the MHD run.

Both Fig.~\ref{fig:entropy_fraction} and \ref{fig:closest_neighbor} indicate that condensation in our simulations is predominantly driven by mixing with pre-existing low-entropy gas rather than by unmediated radiative cooling of the ambient ICM. These findings should be interpreted in the context of our simulation setup: by the time our analysis begins, approximately 1.6 Gyr after the initial single-phase ICM has undergone radiative condensation and the onset of AGN activity, the cluster core has settled into a recurring cycle of cooling and feedback and the exact origin of each cold gas parcel is difficult to establish unambiguously.

\begin{table}[h!]
    \centering
    \caption{Physical quantities tracked for the cooling and non-cooling tracer populations.}
    \label{table:globalquant}
    \begin{tabular}{lccc}
        \hline\hline
        \textbf{Quantity} & \textbf{Unit} & \textbf{Radial Norm.} & \textbf{MHD only} \\
        \hline
        $\rho$                    & $\mathrm{g\,cm^{-3}}$             & \checkmark & --         \\
        $K$                       & $\mathrm{keV\,cm^{2}}$            & \checkmark & --         \\
        $|\nabla\cdot v|$         & $\mathrm{Myr}^{-1}$  & \checkmark         & --         \\
        $|\nabla \times v|$       & $\mathrm{Myr}^{-1}$  & \checkmark         & --         \\
        $t_{\rm cool}/t_{\rm ff}$ & --                    & \checkmark & --         \\
        $\beta^{-1}$  & --                                & \checkmark & \checkmark \\
        \hline
        $\nabla\cdot v$    & $\mathrm{Myr}^{-1}$ & -- & -- \\
        $\mathcal{C}_{\rm comp}$      & -- & -- & -- \\
        $\mathcal{M}_{\rm turb}$  & --                                & --         & --         \\
        \hline
    \end{tabular}
    \tablefoot{Quantities marked \checkmark\ in the radial normalization column are also stored in normalized form, divided by their mean radial profile $\langle \mathcal{Q}(r) \rangle$, alongside their absolute value. The MHD-only column indicates quantities that are unavailable in the purely hydrodynamical run.}
\end{table}

\begin{figure*}[h!]
\includegraphics[width=\textwidth]{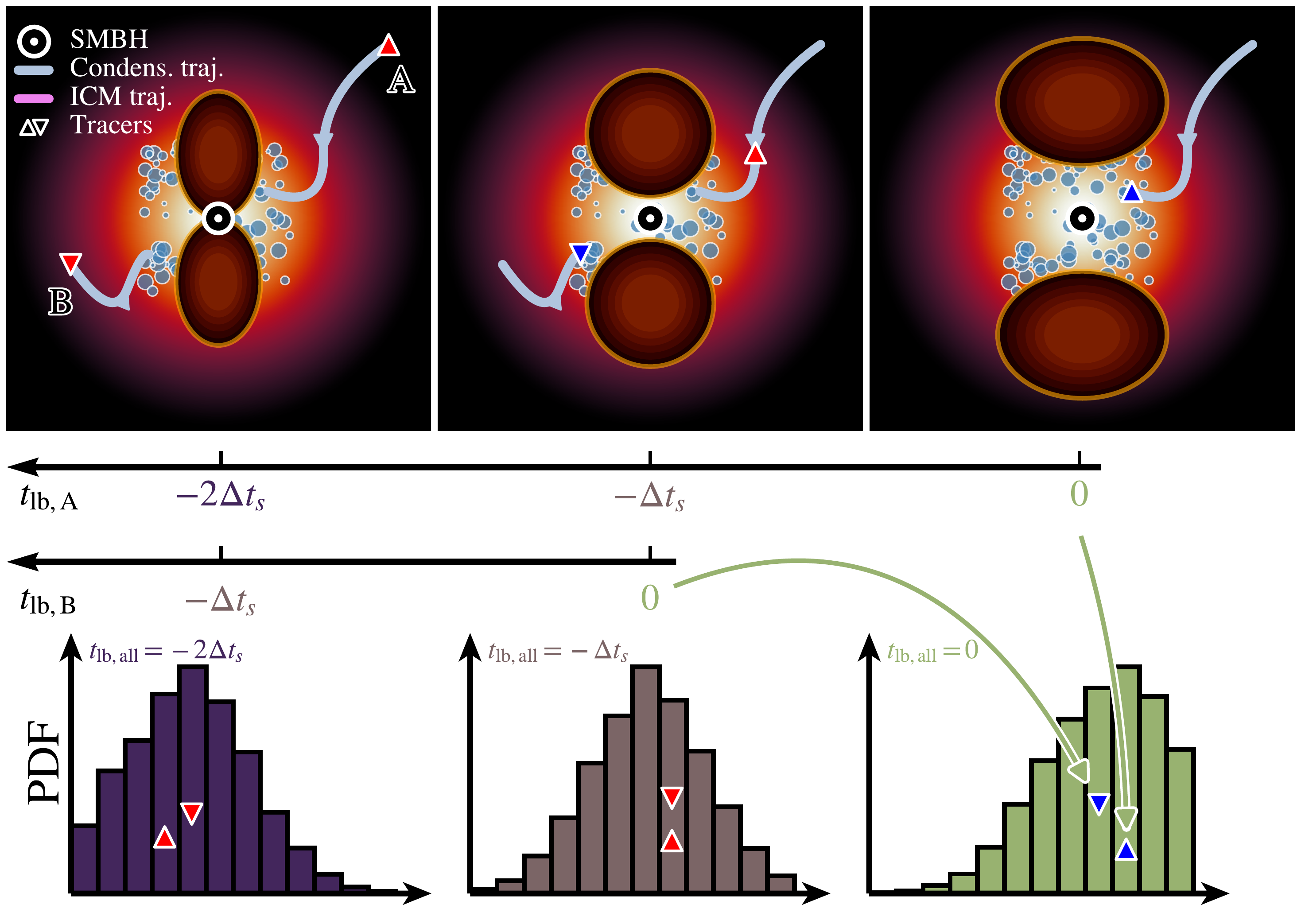}
    \caption{Schematic of the lookback time alignment used in the global tracer analysis. Top: Three consecutive snapshots of the cluster core (separated by a time interval $\Delta t_s$), showing the SMBH, AGN jet lobes (dark red), and ambient cold gas (blue circles). Two representative tracer trajectories (A and B, blue-grey) are overlaid, illustrating gas parcels that condense at different times in the simulation. Filled triangles mark the tracer positions at each snapshot, with color indicating their thermal state: red for hot ($T \geq 5\times10^6$ K) and blue for cold ($T \leq 10^5$ K). The transition between the last red and first blue marker along each trajectory therefore brackets the condensation event. Middle: Each condensation event defines its own lookback time axis, where $t_{\rm lookback} = 0$ corresponds to the first post-condensation snapshot, effectively marking the condensation time given the small $\Delta t_s$. Bottom: At each common lookback time $t_{\rm lb,all}$, contributions from all condensation events are pooled into a single PDF of any given quantity (e.g. density). The rightmost panel shows how individual tracers (filled triangles) map onto the distributions at the first three lookback times ($t_{\rm lb,all} = 0,\ -\Delta t_s,\ -2\Delta t_s$). This alignment enables statistical study of pre-cooling ICM properties across the full tracer population, independent of when individual condensation events occur. Tracking summary statistics of the PDF (e.g. the median) across all available lookback times then reconstructs the temporal evolution of any quantity leading up to condensation, as shown in e.g., Fig.~\ref{fig:full_vorti_comp}.}
    \label{fig:diagram}
\end{figure*}

\subsubsection{Tracer selection and lookback time statistics}
\label{sect:tracerselection}

Having established that the large majority of condensing tracers undergo mixing-driven, abrupt hot-to-cold transitions on timescales shorter than $1\,\rm{Myr}$, we restrict our analysis to this population, excluding the minority that cool more gradually. For each snapshot $s$, we define $\mathcal{P}_{\rm cool}(s)$ as the set of tracers that remain in the hot phase ($T > 5\times10^6\,\mathrm{K}$) over a $\mathcal{W}_{\rm max}=300\,\mathrm{Myr}$ lookback window and reach the cold phase ($T \leq 10^5\,\mathrm{K}$) at time $t_s$; the full $\mathcal{P}_{\rm cool}$ sample is the union over all snapshots. The choice of a 300 Myr lookback window is motivated by the typical cooling time at the radii of interest (a few tens of kpc), ensuring that we capture the full thermodynamic evolution prior to the cooling transition.

\begin{figure*}[h!]
\includegraphics[width=\textwidth]{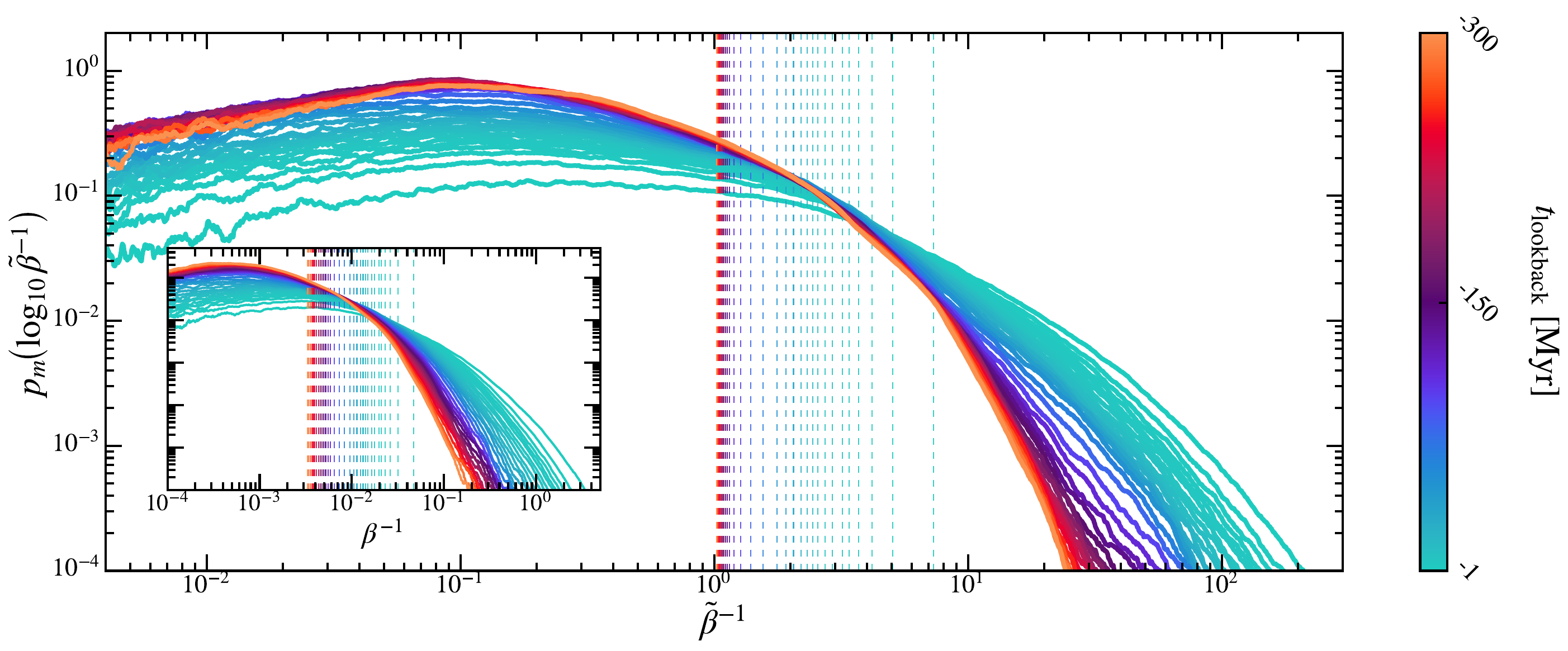}
    \caption{Mass-weighted density PDFs of the inverse plasma beta, $\beta^{-1} \equiv P_{\rm mag}/P_{\rm th}$, for the $\mathcal{P}_{\rm cool}$ population. Each curve corresponds to tracers in the hot phase at a given lookback time prior to their condensation into cold gas. The main panel shows the distribution of $\beta^{-1}$ normalized by the spherically averaged profile of the background ICM, highlighting deviations of the $\mathcal{P}_{\rm cool}$ tracers from the ambient, mostly non-condensing medium. At early lookback times, the distributions closely follow the background profile, indicating that we are probing epochs before the $\mathcal{P}_{\rm cool}$ tracers have diverged from the ambient medium. The inset displays the corresponding distributions of the absolute $\beta^{-1}$. Vertical dashed lines indicate the median value of each distribution at the corresponding lookback time.}
    \label{fig:beta_PDFs}
\end{figure*}

\begin{figure}[h!]
\includegraphics[width=0.5\textwidth]{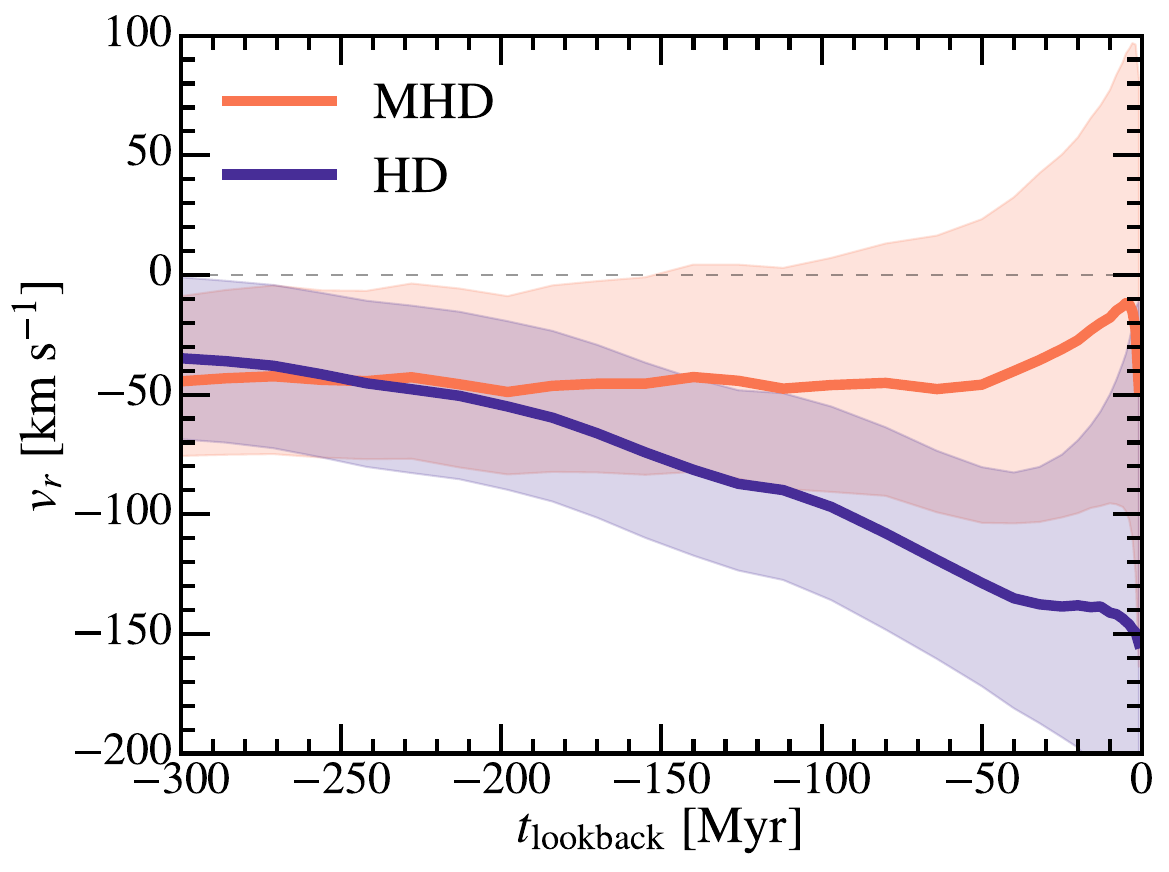}
    \caption{\red{Evolution of the radial velocity of the $\mathcal{P}_{\rm cool}$ tracer population as a function of lookback time, for the MHD (orange) and HD (blue) runs. Shaded bands indicate the temporal variation across snapshots, and solid lines show the global median over all snapshots per lookback-time window.}}
    \label{fig:radvel}
\end{figure}

\begin{figure}[h!]
\includegraphics[width=0.5\textwidth]{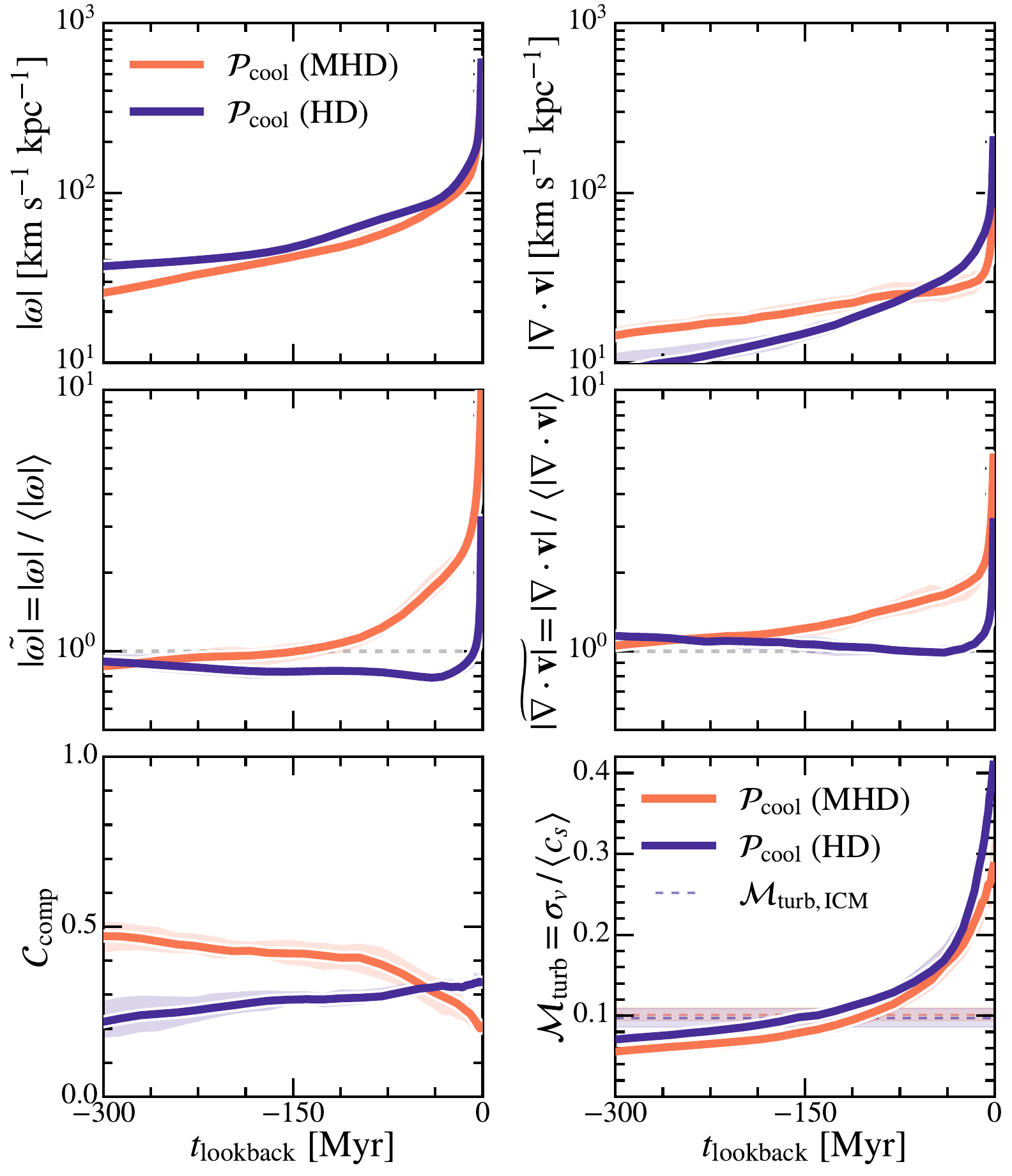}
\caption{Evolution of key kinematic properties of the $\mathcal{P}_{\rm cool}$ tracer population as a function of lookback time, for the MHD (orange) and HD (blue) runs. Top row: median absolute vorticity $|\omega|$ (left) and compressive motions $|\nabla \cdot \mathbf{v}|$ (right). Middle row: the same quantities normalised by the radial profile, $|\tilde{\omega}|$ and $|\widetilde{\nabla \cdot \mathbf{v}}|$, quantifying deviations relative to the background ICM (grey dashed line: $\vert \tilde{\omega}\vert = 1$). Bottom row: median compressive fraction $C_{\rm comp}$ (left) and turbulent Mach number $\mathcal{M}_{\rm turb}$ (right); shaded bands and dashed lines in the Mach number panel show the median and time variation of the background ICM for reference. In all panels, shaded bands indicate the temporal variation across snapshots and solid lines the global median over all snapshots per lookback-time window.}
\label{fig:full_vorti_comp}
\end{figure}

Each tracer in $\mathcal{P}_{\rm cool}$ is assigned a lookback time axis defined such that $t_{\rm lookback} = 0$ coincides with its condensation event; aligning all such axes and pooling tracer states at each common lookback time yields a PDF of physical quantities of interest (e.g., entropy). Tracking summary statistics of these PDFs (e.g., the median) across lookback times then reconstructs the temporal evolution of that quantity in the lead-up to condensation. This procedure is illustrated in Fig.~\ref{fig:diagram}. Since most tracked quantities exhibit a pronounced radial dependence, we normalize each tracer value by the mean radial profile $\langle \mathcal{Q}(r,t) \rangle$ computed from all hot-phase cells; normalized quantities are indicated by a tilde (e.g., $\tilde{K} \equiv K/\langle K(r) \rangle$). Quantities with a weaker radial dependence are retained in absolute form only. This normalization isolates the deviation of cooling gas from the hot background and quantifies both the timescale and amplitude of that divergence. The full set of tracked quantities is summarized in Table~\ref{table:globalquant}.

We define $\mathcal{C}_{\rm comp}$ as the local compressive fraction of the local velocity gradient, defined as:

\begin{equation}
    \mathcal{C}_{\rm comp} = \frac{|\nabla \cdot v|}{\,\sqrt{(\nabla \cdot v)^2 + (\nabla \times v)^2}},
\end{equation}

ranging from 0 for purely solenoidal flow to 1 for purely compressive flow. We also compute the turbulent Mach number of the tracers by loading all gas cells within a sphere of radius $r = 5\,\mathrm{kpc}$\footnote{We have verified that adopting alternative radii leaves our results qualitatively unchanged.} centered on the tracer position. We then define

\begin{equation}
    \mathcal{M}_{\rm turb} = \frac{\sigma_v}{\langle c_s \rangle},
\end{equation}

where $\sigma_v$ is the velocity dispersion and $\langle c_s \rangle$ is the mean sound speed, both evaluated within the sphere.

\subsubsection{Thermodynamic and dynamical evolution before cooling}
\label{sect:thermoevol}

As an illustration of the tracer analysis described in Sect.~\ref{sect:tracerselection}, Fig.~\ref{fig:beta_PDFs} shows the mass-weighted distributions of the inverse plasma beta, $\beta^{-1}$ (hereafter referred to as magnetization), for 32 lookback times up to $-300$ Myr for the cooling tracer population, $\mathcal{P}_{\rm cool}$ in our MHD fiducial run. The main panel shows the distribution of $\tilde{\beta}^{-1} \equiv \beta^{-1}/\langle \beta^{-1} \rangle$, where the normalization is taken with respect to the spherically averaged background ICM profile, highlighting deviations of the $\mathcal{P}_{\rm cool}$ tracers from the ambient, mostly non-condensing medium. The inset shows the corresponding distributions of the absolute $\beta^{-1}$. At early lookback times, the distributions closely follow the background profile, confirming that we are probing epochs before the $\mathcal{P}_{\rm cool}$ tracers have significantly diverged from the ambient medium.  As the lookback time decreases, the distributions gradually shift toward higher magnetization, with median values reaching an order of magnitude above their initial levels in both the normalized and absolute distributions. In the following analysis, we focus on the time evolution of the median values of these mass-weighted profiles and examine how the different quantities diverge relative to their initial state at a lookback time of $-300$ Myr.

\red{Tracers in the $\mathcal{P}_{\rm cool}$ population are on average characterized by distinct evolutionary pathways in terms of radial trajectory, as visible in Fig.~\ref{fig:radvel}. Consistent with the trajectories visualized in Fig.~\ref{fig:mainfig}, tracers in the pure hydro run preferentially follow infalling trajectories, with infall speeds steadily increasing to $\sim 150$ km s$^{-1}$ in the last few Myr before condensation. Conversely, condensing tracers in the MHD run are characterized by radial velocities closer to -50 km s$^{-1}$, with roughly half of the tracers with positive radial velocities in the last few Myr before condensation. This supports a picture in which condensing gas parcels follow distinct pathways in pure hydro and MHD, i.e. mixing with the massive central cold disk in pure hydro and mixing with uplifted clumps in the MHD case.}

\begin{figure}[h!]
\includegraphics[width=0.5\textwidth]{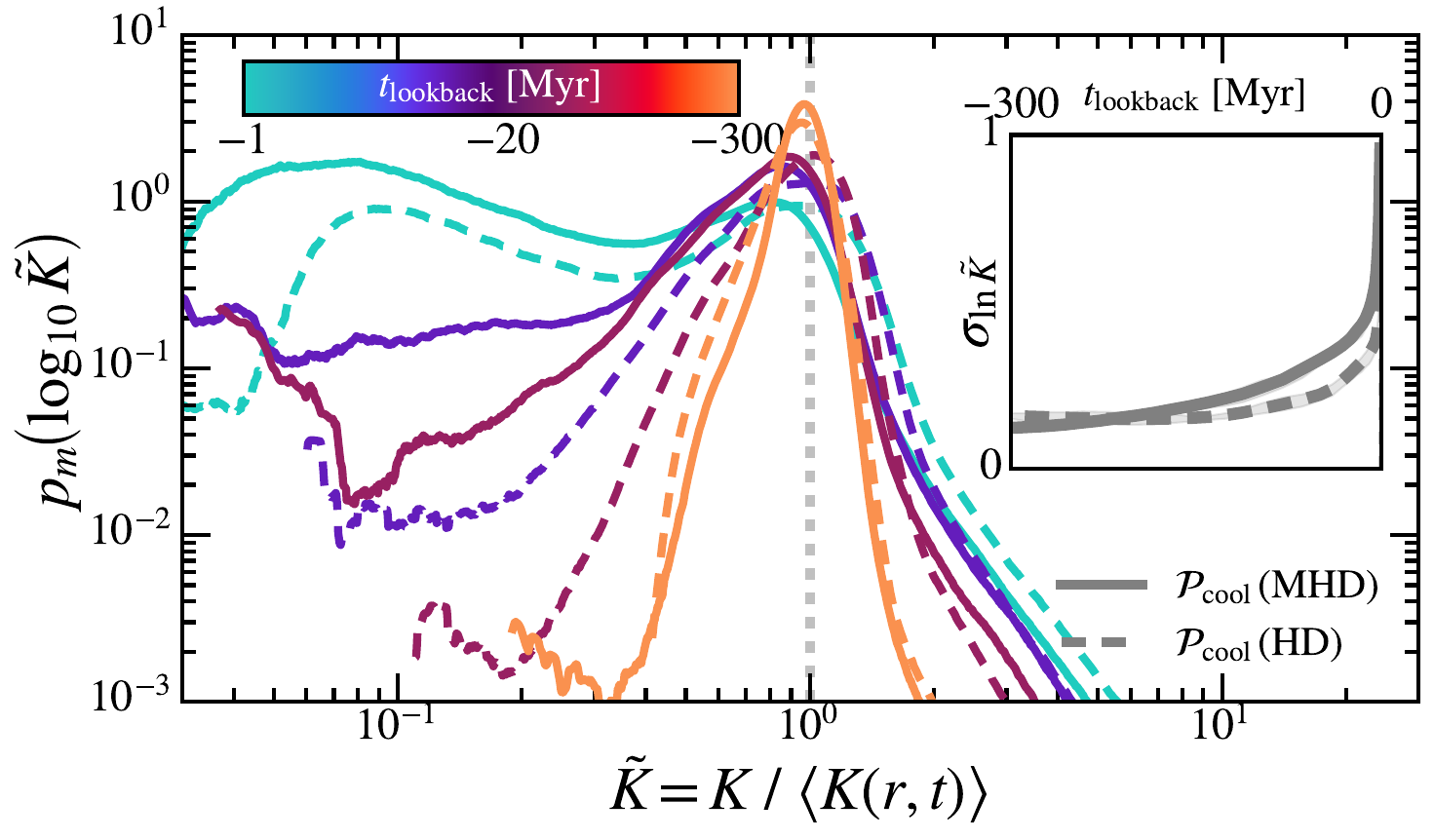}
    \caption{Mass-weighted probability distribution of the normalized entropy, $p_m(\log_{10}\tilde{K})$, where $\tilde{K} = K/\langle K(r,t)\rangle$ is the entropy of each tracer normalized to the local radial mean, for the $\mathcal{P}_{\rm cool}$ tracer population. Solid curves correspond to the MHD run; dashed curves indicate the purely hydrodynamic (HD) case. Colors encode lookback time on a logarithmic scale prior to the cooling event, ranging from $\sim$300~Myr (orange-red) to $\sim$1~Myr (cyan). The vertical dotted line marks $\tilde{K} = 1$, i.e., alignment with the background distribution. The inset figure shows the corresponding evolution of the logarithmic entropy width $\sigma_{\ln\tilde{K}}$ as a function of lookback time for both the MHD (solid) and HD (dashed) runs. This behaviour is qualitatively consistent with the schematic picture of entropy fluctuation growth depicted in Fig.~3 of \citet{Voit_2021}.}
    \label{fig:entropyfluctuations}
\end{figure}

In Fig.~\ref{fig:full_vorti_comp}, we show the evolution of the absolute vorticity $\vert \omega \vert$, absolute compression $\vert \nabla \cdot \mathbf{v} \vert$, compressive fraction $\mathcal{C}_{\rm comp}$, and turbulent Mach number $\mathcal{M}_{\rm turb}$ for the $\mathcal{P}_{\rm cool}$ population in both our fiducial MHD and purely hydrodynamic runs. Although both runs exhibit somewhat similar absolute values for vorticity and compression at fixed lookback time \red{(consistent with the slightly higher background vorticities in the pure HD run and the comparable compression ICM profiles shown in Appendix~\ref{app:additionaldiag})}, a key difference concerns the lookback time at which the $\mathcal{P}_{\rm cool}$ population begins to deviate from the background ICM profile. In MHD, both the normalized vorticity $\widetilde{|\omega|}$ and normalized compression $\widetilde{|\nabla \cdot \mathbf{v}|}$ start diverging from unity as early as $\vert t_{\rm lookback} \vert \sim 150$~Myr, indicating that the gas destined to cool is kinematically distinguished from the non-condensing background long before condensation occurs. In contrast, the pure hydro run shows no significant departure until $\vert t_{\rm lookback} \vert \lesssim 30$~Myr, suggesting a much more abrupt and localized build-up of kinematic activity immediately prior to condensation. The evolution of $\mathcal{C}_{\rm comp}$ reveals a further qualitative difference: in the hydro run it increases monotonically as cooling approaches, indicating progressively dominant compressive motions consistently with \citet{Sotira_2026}, whereas in MHD it remains roughly constant before falling below the hydro value at $\vert t_{\rm lookback} \vert \lesssim 50$~Myr, suggesting that vortical motions grow in relative importance as condensation approaches, or that magnetic pressure is suppressing compressive flows.

The turbulent Mach number $\mathcal{M}_{\rm turb} = \sigma_v / \langle c_s \rangle$ increases monotonically in both runs, rising from $\mathcal{M}_{\rm turb} \sim 0.1$ at $\vert t_{\rm lookback}\vert \sim 200$~Myr to $\sim 0.3$--$0.4$ at $t_{\rm lookback} \sim 0$, confirming that cooling gas is embedded in a progressively more turbulent environment. The two runs track each other closely at early times but diverge at $\vert t_{\rm lookback}\vert \lesssim 25$~Myr, where the hydro run reaches systematically higher Mach numbers ($\mathcal{M}_{\rm turb} \approx 0.4$ versus $\approx 0.28$ in MHD at the cooling transition), suggesting that in the absence of magnetic fields more turbulent driving is required for condensation \citep[see e.g.,][]{Wibking_2024}, although this interpretation may be affected by selection effects and does not necessarily imply a direct causal role of magnetic fields (see associated discussion in Sect.~\ref{sect:limits}). Taken together, these results point to a qualitatively different condensation pathway in the magnetized run, in which increased levels of vortical motion play a more central role than compressive flows.

In Fig.~\ref{fig:entropyfluctuations}, we show the distribution of the normalized entropy, $\tilde{K}\equiv K \ / \ \langle K(r,t) \rangle$, for the $\mathcal{P}_{\rm cool}$ tracer particle population. The orange, maroon and purple and cyan curves correspond to the distributions for the MHD case at 200, 100, 15 and 1 Myr lookback time (respectively). The dashed lines are the corresponding distributions for the purely hydrodynamical case. The $\mathcal{P}_{\rm cool}$ population is largely consistent with the background ICM profile at $\vert t_{\rm lookback}\vert=300$ Myr, with the distribution centered on $K/\langle K \rangle = 1$. At later times (starting from $\vert t_{\rm lookback}\vert \sim 100$ Myr), the distribution broadens and exhibits a raised lower-end component. At both $\vert t_{\rm lookback} \vert = 15$ and $1$~Myr, the amplitude of the low-$\tilde{K}$ component is systematically higher in the MHD case than in the purely hydrodynamical run by roughly an order of magnitude at $\tilde{K} \sim 0.1$. The two runs exhibit comparable distributions on the high-entropy side ($\tilde{K} > 1$), indicating that the difference is localized entirely to the sub-median, low-entropy tail. This asymmetry shows that entropy fluctuations grow well before the onset of mixing.

\begin{figure}[h!]
\includegraphics[width=0.5\textwidth]{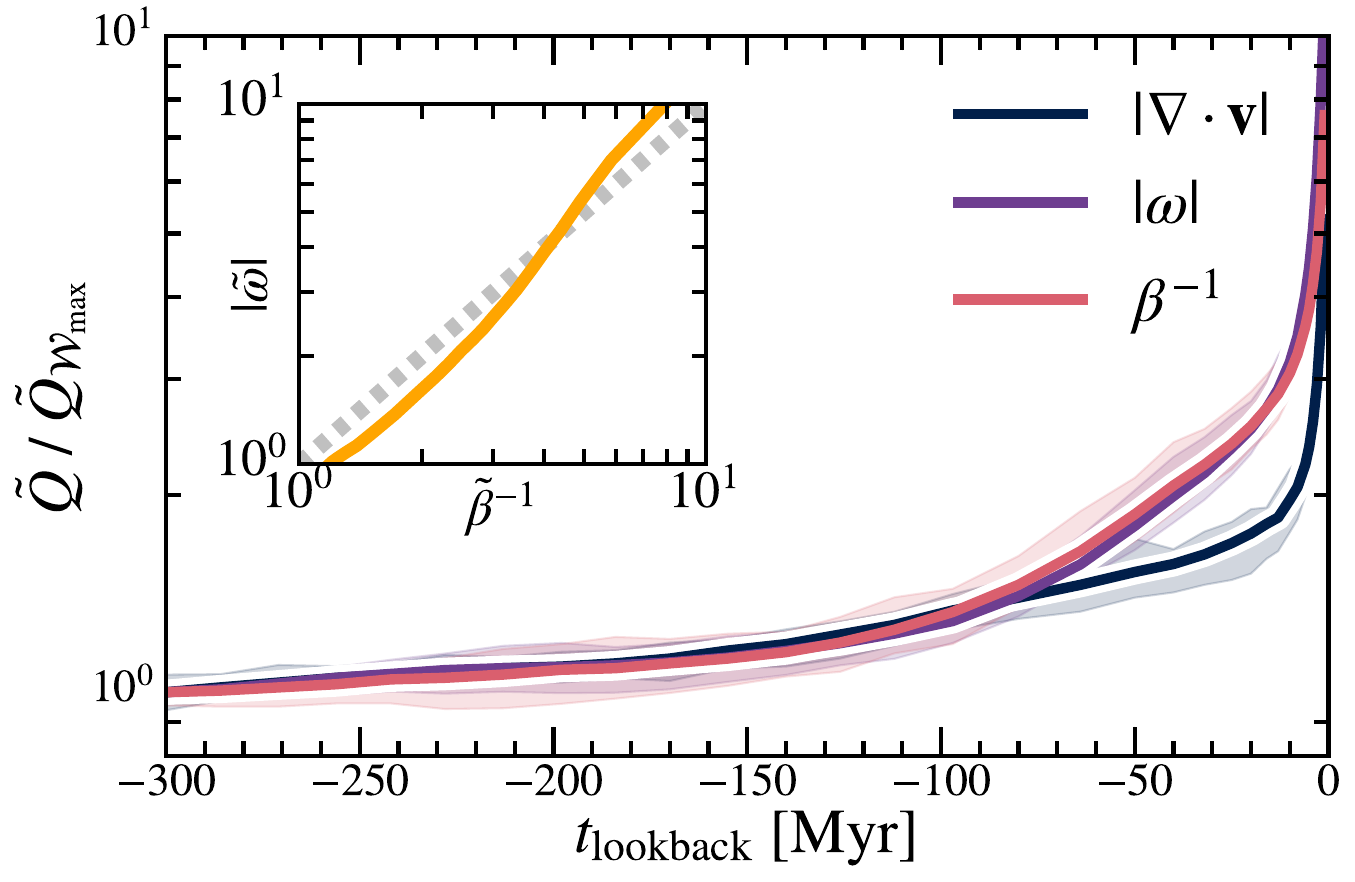}
    \caption{Evolution of the mean local compression $|\nabla \cdot \mathbf{v}|$, vorticity $|\omega|$, and inverse plasma beta $\beta^{-1}$ for the $\mathcal{P}_{\rm cool}$ population in our MHD run. Each quantity $\mathcal{Q}$ is normalized by its spherically averaged radial profile $\langle \mathcal{Q}(r,t) \rangle$ to isolate deviations from the ambient non-condensing gas, and further divided by its value at $\vert t_{\rm lookback} \vert = 300$ Myr so that all curves begin at unity, highlighting their co-evolution. Shaded bands indicate snapshot-to-snapshot variation within each lookback time bin. The inset shows the absolute correlation between $|\tilde{\omega}|$ and $\tilde{\beta}^{-1}$, with the dashed grey line indicating a linear scaling for reference.}
    \label{fig:vorticity_vs_mag}
\end{figure}

\begin{figure*}[h!]
\includegraphics[width=\textwidth]{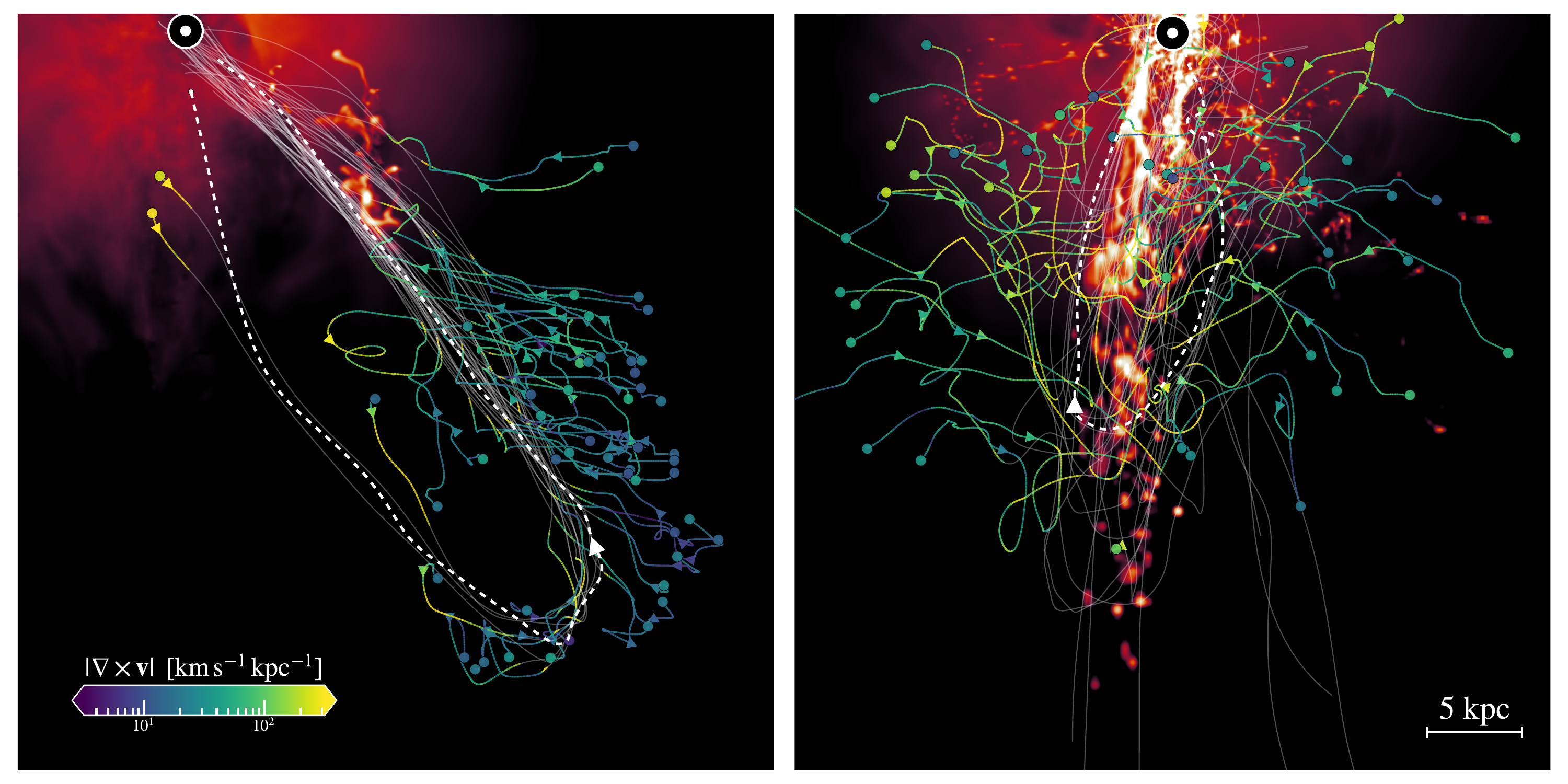}
    \caption{Tracer trajectories from the $\mathcal{P}_{\rm cool}$ population overlaid on density projections of cold gas, illustrating the formation pathways of two distinct cold structures encountered in the MHD run: an isolated ballistic cloud (BC\#1, left) and an extended filament (right). The thick dashed line traces the trajectory of each structure's main progenitor, revealing an initial uplift phase followed by subsequent infall. Colored lines represent individual tracer particles that begin hot and eventually cool to join the cold structure. The color encodes the vorticity magnitude, transitioning to grey once each tracer cools to the temperature floor.}
    \label{fig:tracer_closeup}
\end{figure*}

In Fig.~\ref{fig:vorticity_vs_mag}, we focus on our MHD run and show the evolution of the normalized compression $\vert \nabla \cdot \mathbf{v} \vert$, vorticity $\vert \boldsymbol{\omega} \vert$, and inverse plasma beta $\beta^{-1}$ for the $\mathcal{P}_{\rm cool}$ population. While all three quantities increase as the lookback time decreases, normalized vorticity and inverse plasma beta exhibit a remarkably tight co-evolution over the entire time interval. This correspondence primarily reflects a coupling in their deviations from the local background, rather than a strict point-wise scaling between the two absolute quantities. Both quantities show a gradual rise until $\vert t_{\rm lookback} \vert \sim 100$~Myr, after which their slopes become steeper, marking a transition in the amplification rate of vorticity and magnetic energy. Around $\vert t_{\rm lookback} \vert \sim 10$~Myr, a further steepening occurs, with all three quantities increasing sharply, consistent with strong compression of the hot plasma shortly before condensation. We have verified independently that the magnetic energy growth rate $\Gamma$ scales with enstrophy magnitude, $\log_{10}(\Gamma) \propto \log_{10}(|\boldsymbol{\omega}|^2)$, consistent with previous simulations of small-scale dynamo amplification in a multiphase medium \citep{Gent_2023}, suggesting that the enhanced magnetic field strength may result from increased turbulence, potentially driven by AGN activity or the motion of nearby low-entropy gas. This connection supports a picture in which turbulent motions amplify magnetic fields prior to condensation. In turn, the strengthened magnetic field may further influence the thermodynamic evolution of the gas by modifying its stability and mixing properties.

\subsection{Evolution of individual cold structures}

In this subsection, we focus on our fiducial MHD run and study the evolution of individual substructures identified in our simulation after the injection of tracers. Two distinct kinds of structures are identified: (i) massive filaments forming at the edge and in the wake of the rising cavities, and (ii) individual clouds, initially entrained by the jets, and reaching much higher altitudes (up to $\sim 70$ kpc away from the SMBH). This second category is particularly interesting because these clouds spend most of their life cycle evolving in relatively quiescent regions of the ICM, away from jet activity or any perturbations related to the AGN activity. Their trajectory is mostly ballistic and represents a clean laboratory to study the magnetic properties of the cold gas in the ICM, and potential effects of tension on its kinematics, such as predicted in \citet{Voit_2026}. We identify two of these structures, which we label “ballistic clouds” (BC\#1 and BC\#2), and study their evolution from their initial coupling with the jet, until their accretion onto the AGN on their way back towards the cluster center. In Sect.~\ref{sect:individualstruc}, we compare the lookback histories of tracers in individual structures with the global analysis presented in Sect.~\ref{sect:globaltracers}. We then examine the dynamics of the cold phase in Sect.~\ref{sect:ballistic}.

For each of these three particular structures, we use the tracer particles to efficiently reconstruct the accretion story across snapshots. Initially, we start by identifying all the cells belonging to a given structure (i.e., the filament and BCs) at a reference snapshot and time $t_{\rm{ref}}$ where the structure can be cleanly separated from the background using a clump finding algorithm. By tracing these populations back through previous snapshots, we reconstruct the history of the clumps, following tracers from the hot phase as they cool and accrete onto the forming structures. Sample tracer trajectories for gas accreted by BC\#1 and the filament  are shown in Fig.~\ref{fig:tracer_closeup}. A notable qualitative  difference emerges between the two structures: the filament draws in  gas parcels from a broad region of the ICM, with a sphere of influence  of roughly 40~kpc, whereas the ballistic cloud accretes gas almost  exclusively along a narrow channel aligned with its trajectory.

\begin{figure}[h!]
\includegraphics[width=0.5\textwidth]{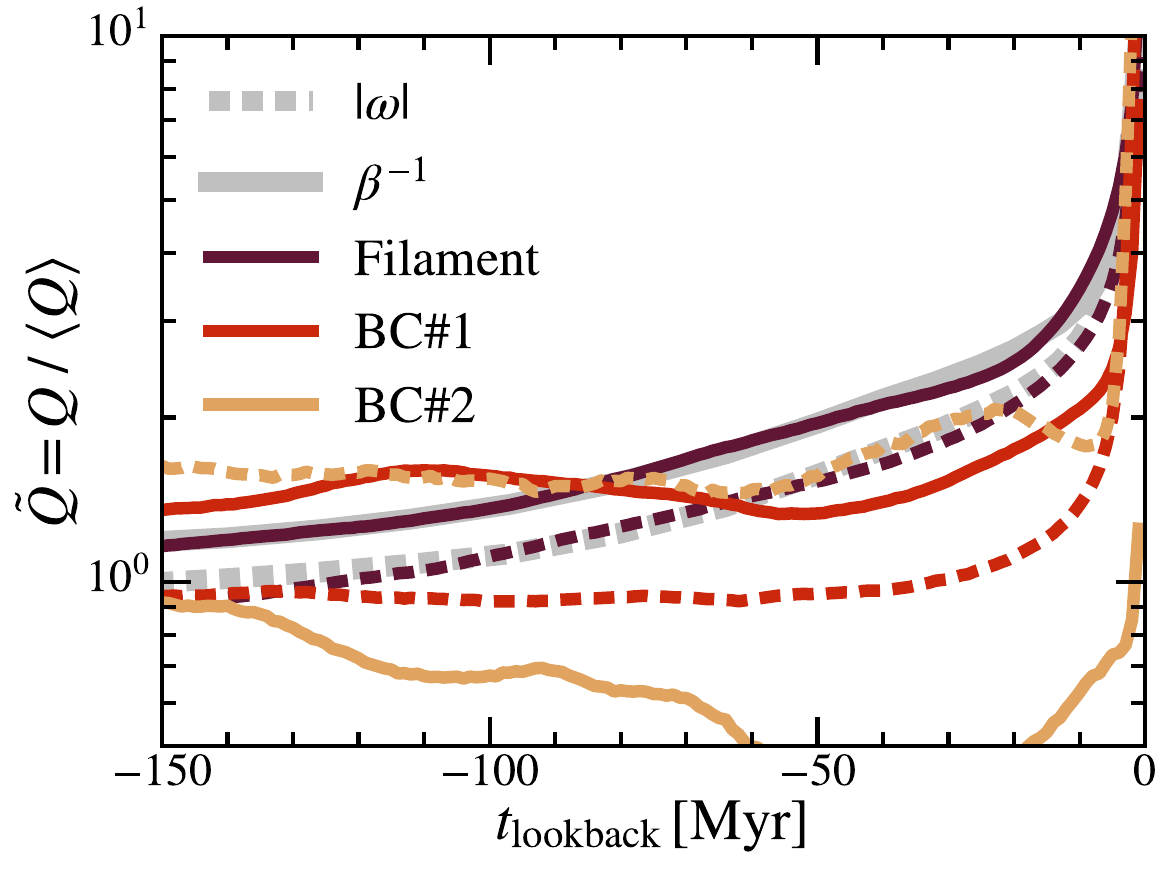}
    \caption{Vorticity and inverse $\beta$ as a function of time before cooling for the three individual cold structures identified in the MHD run (coloured lines). The thick grey lines reproduce the global $\mathcal{P}_\mathrm{cool}$ statistics from Fig.~\ref{fig:vorticity_vs_mag} for reference (dashed: vorticity; solid: inverse $\beta$). The massive filament tracks the global population closely, whereas the two ballistic clouds (BC) show a markedly different evolution.}
    \label{fig:vorticity_mag_filcloud}
\end{figure}

\begin{figure*}[h!]
\includegraphics[width=\textwidth]{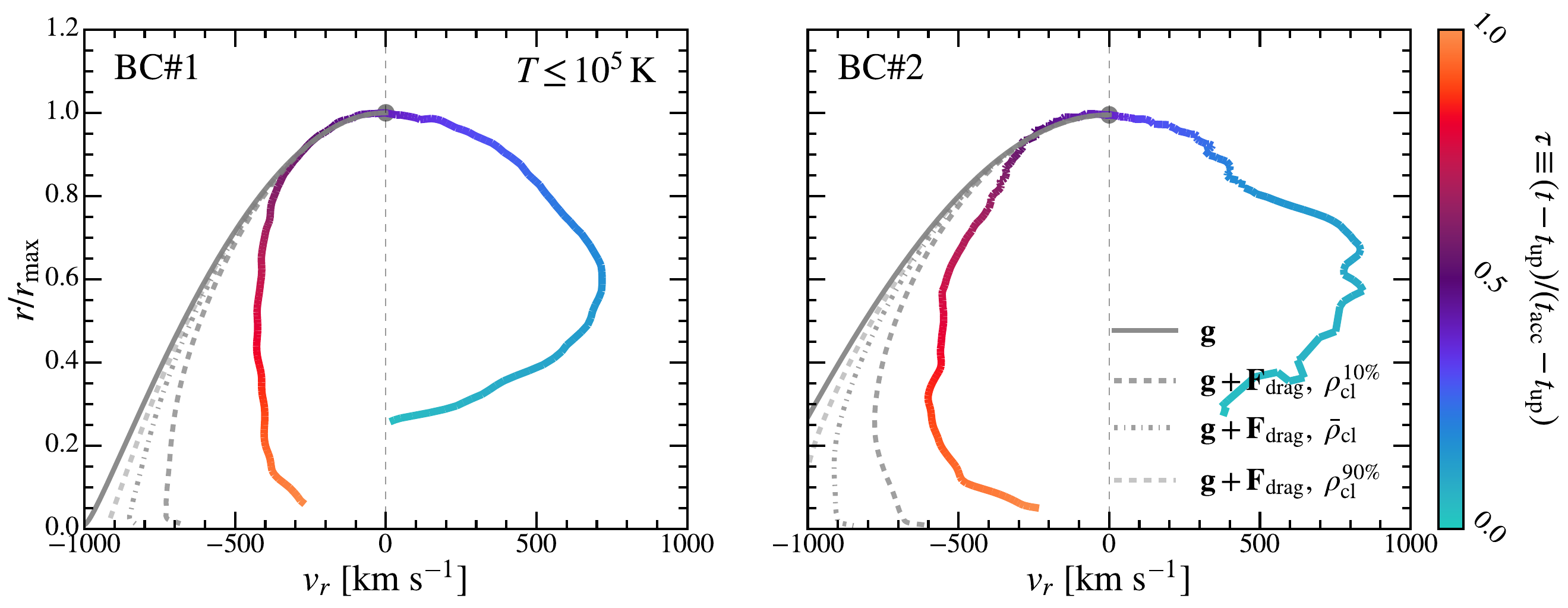}
    \caption{Trajectories in the radial distance–radial velocity plane are shown for the two individual clouds identified in our MHD simulation. Both clumps start with an initial outward radial velocity resulting from their coupling with the AGN jet. Overlaid are trajectories predicted by a basic semi-implicit Euler integration including gravity alone (solid grey line) and gravity plus ram pressure drag for three choices of cloud density: the 10th and 90th percentiles of the mass-weighted density PDF of the cloud's cold gas ($\rho_{\rm cl}^{10\%} = 3\times10^{-24}\ \rm g\,cm^{-3}$ and $\rho_{\rm cl}^{90\%} = 3\times10^{-23}\ \rm g\,cm^{-3}$, dashed lines), and the mass-weighted mean density $\bar{\rho}_{\rm cl} = 1\times10^{-23}\ \rm g\,cm^{-3}$ (dash-dotted line). The cloud cross-section is derived from its time-dependent total mass and assumed density. For all density assumptions, ram pressure drag alone is insufficient to account for the observed reduction in the cloud's terminal velocity. The maximum altitude reached by BC\#1 and \#2 are 45 and 70 kpc, respectively. Trajectories are colour-coded by the rescaled time $\tau \equiv (t - t_{\rm up})/ (t_{\rm acc} - t_{\rm up}) \in [0, 1]$, where $\tau = 0$ and $\tau = 1$ mark the onset of uplift and accretion, respectively.}
    \label{fig:drag}
\end{figure*}

\begin{figure*}[h!]
\includegraphics[width=\textwidth]{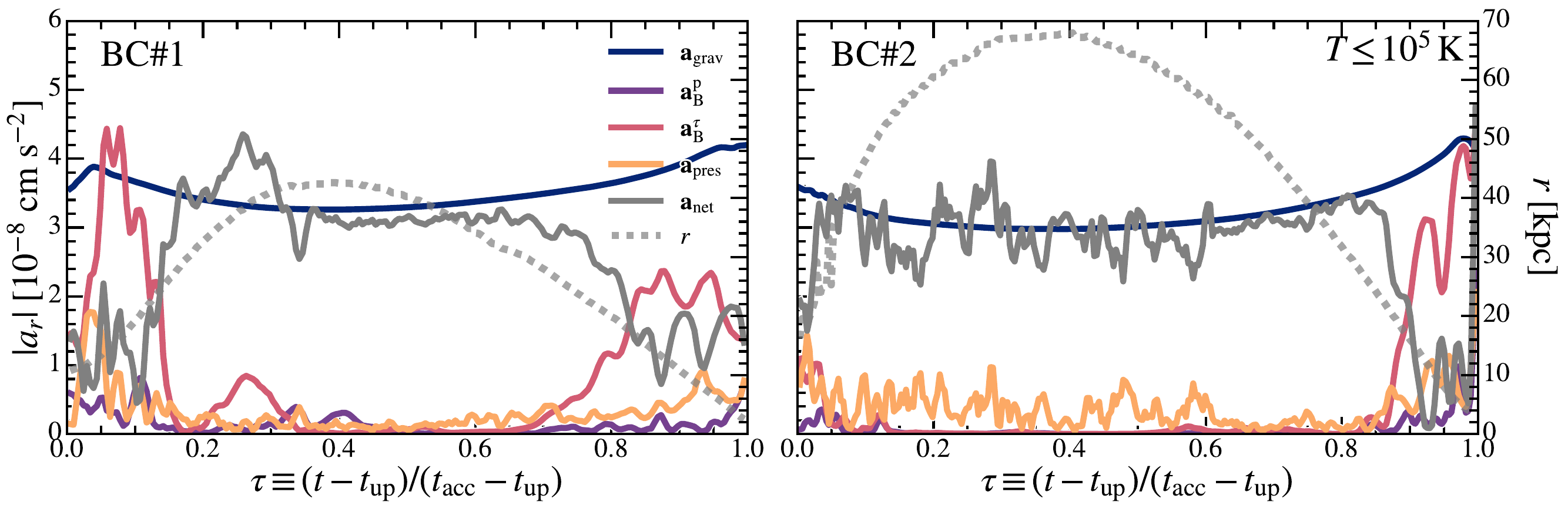}
    \caption{Norms of the radial components of the four acceleration terms for the two isolated clouds (left and right panels). Each term is computed as the mass-weighted contribution from all cells within the cloud. The net acceleration is shown as the solid grey line, while the dashed grey line indicates the cloud altitude. \red{Acceleration terms appearing close to zero are dynamically negligible at that epoch and do not contribute meaningfully to the cloud evolution}. In both clouds, the magnetic tension gradually increases during infall, partially counteracting gravity and reaching up to $\sim 100\%$ of the gravitational acceleration in the second cloud.}
    \label{fig:acceleration}
\end{figure*}

\subsubsection{Cooling pathways}
\label{sect:individualstruc}

Fig.~\ref{fig:vorticity_mag_filcloud} shows the evolution of the radially normalized vorticity, $|\tilde{\omega}|$, and $\tilde{\beta}^{-1}$ as a function of time before cooling for the three individual cold structures, alongside the global $\mathcal{P}_\mathrm{cool}$ statistics reproduced for reference. The massive filament tracks the global population with striking fidelity across the entire 150 Myr window, both in its vorticity and magnetization evolution. This close agreement reflects the fact that filamentary structures of this kind represent the dominant population among cooling tracers in the ICM, and that the global statistics presented in Sect.~\ref{sect:globaltracers} are largely shaped by their collective behavior. Notably, the global tracer analysis in Sect.~\ref{sect:globaltracers} was restricted to gas that cooled at least 300 Myr after the beginning of the tracer-active phase of the simulation, in order to retain the full lookback history. The massive filament, which formed only $\sim$150 Myr after the start of this phase, is therefore not included in that sample, yet its evolution is fully consistent with the global statistics, hinting at a universal condensation pathway for filament-like structures.

The two ballistic clouds tell a markedly different story. Because they spend most of their evolution at significantly larger cluster-centric radii (up to $r \sim 70$ kpc, compared to $\sim$25--30 kpc for the filament), they inhabit a much more weakly magnetized environment. The quantities plotted here are radially normalized, meaning the actual magnetic field strength of the BCs at the top of their trajectories is substantially lower than that of the filament, and for BC\#2 it may fall below the radial mean field strength profile altogether (see Appendix~\ref{fig:B_norm}). At such low ambient magnetization, the pre-cooling evolution is expected to more closely resemble what one would observe in a purely hydrodynamical setup. Consistent with this expectation, both BC\#1 and BC\#2 show weaker and more delayed growth in both vorticity and inverse $\beta$, with little coherent evolution until the final $\sim$50 Myr before cooling. This behaviour supports a picture in which the coupled vorticity--field amplification seen in the global population and the filament is driven primarily by the stronger magnetic environment at lower cluster-centric radii, and is largely absent for clouds condensing in the more quiescent, weakly magnetized outskirts of the core.

\subsubsection{Kinematics of the ballistic clouds}
\label{sect:ballistic}

In Fig.~\ref{fig:drag}, we show the phase-space trajectory of each BC in the $(v_r,\, r/r_{\rm max})$ plane, where $r_{\rm max}$ is the peak altitude reached by the clump. Each trajectory is colour-coded by the rescaled time $\tau \equiv (t - t_{\rm up})/(t_{\rm acc} - t_{\rm up})$, where $t_{\rm up}$ is the time at which the cloud begins its uplift phase and $t_{\rm acc}$ is the time at which it is accreted, so that $\tau = 0$ and $\tau = 1$ mark the beginning and end of the cloud's journey, respectively. After an initial phase of positive radial velocity, both BCs reach a maximum altitude and fall back towards the cluster centre. Notably, the radial velocity of both BCs reaches a maximum absolute value of $\sim 500\ \rm km\,s^{-1}$ during infall, well below the free-fall expectation, suggesting that an additional deceleration mechanism is at play. A natural candidate is ram pressure drag arising from the interaction of the infalling cloud with the ambient ICM. To test this hypothesis, we develop a simple analytical model to estimate the expected trajectory.

Starting from the observed turnaround point with zero radial velocity, we integrate the radial equation of motion using a symplectic, semi-implicit Euler scheme. We consider two cases: gravity alone, and gravity plus ram-pressure drag. The radial projection of the drag force takes the form:

\begin{equation}
    F_{\rm{drag}, \mathit{r}} = -\frac{1}{2} C_D \rho_{\rm ICM}(r) A_{\rm cl} v_r |v_r|,
\end{equation}
where $C_D = 1$ is the drag coefficient, $\rho_{\rm ICM}(r)$ is the ICM density evaluated from mass-weighted radial profiles at the instantaneous cloud position, and $A_{\rm cl} = \pi R_{\rm cl}^2$ is the cross-sectional area of the cloud inferred from its time-evolving mass $M_{\rm cl}(t)$ assuming spherical geometry and a fixed internal density, $\rho_{\rm cl}$, such that $R_{\rm cl} = \left(3 M_{\rm cl} / 4\pi\rho_{\rm cl}\right)^{1/3}$. The mass evolution $M_{\rm cl}(t)$ is directly taken and interpolated from the actual cloud tracked in the simulation. While this is not strictly rigorous, as different ballistic trajectories could in reality lead to different accretion histories, it provides a good estimate of the mass variation throughout the infall phase. We have verified that fixing the cloud mass to any typical value within the relevant range does not significantly affect the result. To bracket the plausible range of bulk cloud densities, we compute the mass-weighted density PDF of the cold gas ($T < 10^5$ K) within the cloud, and adopt the 10th percentile ($\rho_{\rm cl}^{10\%} = 3\times10^{-24}\ \rm g\,cm^{-3}$), the mass-weighted mean ($\bar{\rho}_{\rm cl} = 1\times10^{-23}\ \rm g\,cm^{-3}$), and the 90th percentile ($\rho_{\rm cl}^{90\%} = 3\times10^{-23}\ \rm g\,cm^{-3}$) as representative values. A lower assumed density yields a larger cross section and therefore stronger drag, while a higher density compresses the cloud into a smaller, less-decelerated body. 

As evident for both clouds in Fig.~\ref{fig:drag}, none of the four analytic curves accurately reproduce the measured radial velocity evolution. Even when adopting the lower estimate of the cloud density, the predicted reduction in radial velocity remains roughly a factor of two smaller than observed in our simulation. Given the actual geometry of these two BCs, significantly elongated along the radial direction, and potential mitigating effect of magnetic draping, even these analytic estimates of ram pressure drag are likely overestimates.

\begin{figure}[h!]
\includegraphics[width=0.5\textwidth]{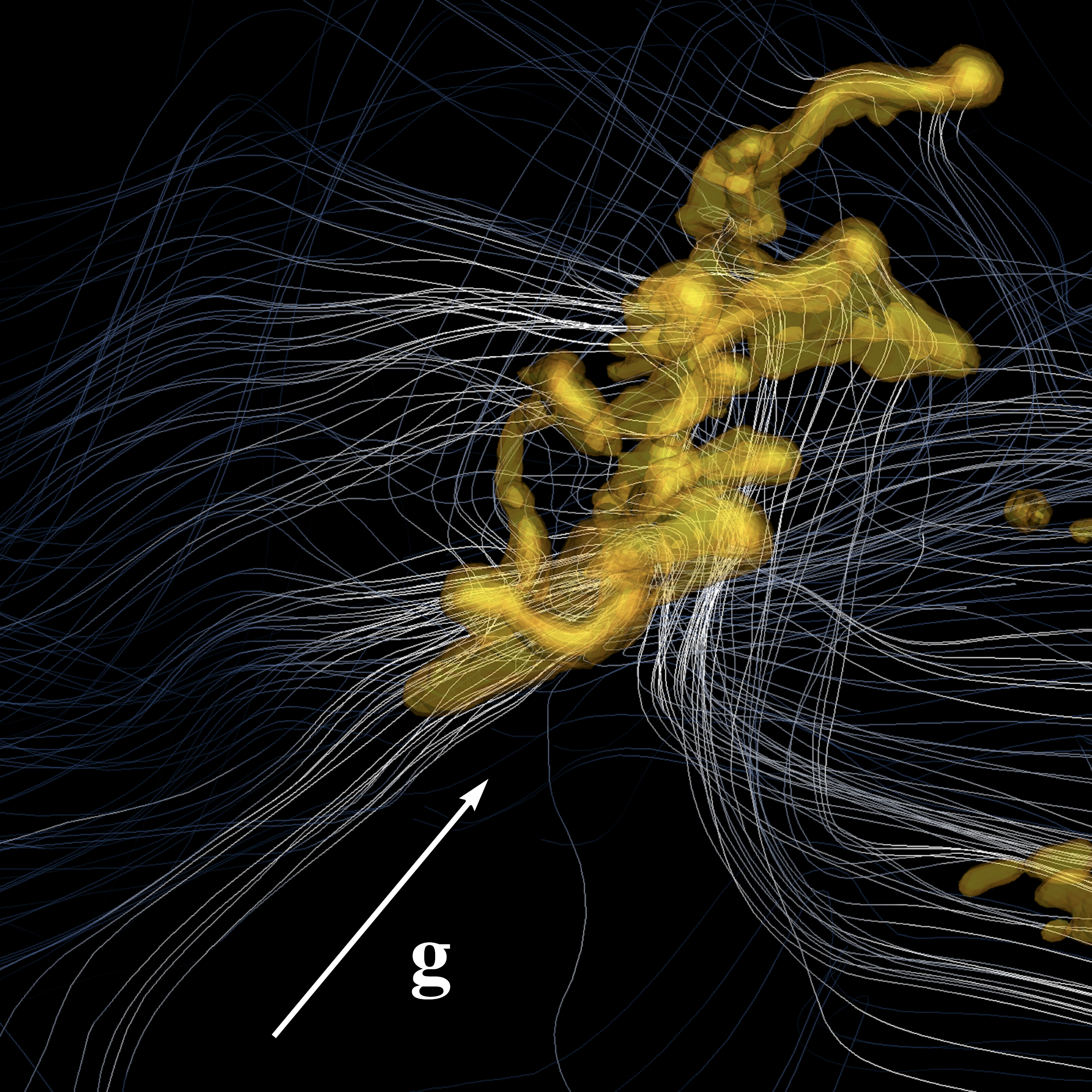}
    \caption{Perspective volume rendering of the $T = 10^{5}\,\mathrm{K}$ isothermal surface of infalling cold cloud \#1 (orange surface), along with magnetic field lines being dragged by the flow and exhibiting the characteristic “drip” morphology described in \citet{Voit_2026}. The width of the image is approximately 15 kpc. The arrow indicates the direction of the local gravitational acceleration vector $\mathbf{g}$.}
    \label{fig:HVC_mag}
\end{figure}

To assess the contributions of different forces to the cloud's net radial acceleration, we compute the total acceleration associated with each of the four terms (gravity, thermal pressure, magnetic pressure, and magnetic tension) and project them along the radial direction. The results for both BCs are shown in Fig.~\ref{fig:acceleration}, where all terms are plotted as absolute values to facilitate comparison of their magnitudes. For most snapshots, gravity points toward the cluster center, while the other terms act outward. The net acceleration is indicated by the solid grey line, and the cloud's altitude is shown with a grey dashed line. Notably, magnetic tension becomes the dominant force opposing gravity during the final 20–30 kpc of the BCs’ infall. For BC\#2, the combined effect of magnetic tension and thermal pressure is sufficient to fully counteract gravity. Fig.~\ref{fig:HVC_mag} provides a visual impression of this dynamical regime. The perspective volume rendering shows BC\#1 (orange surface) together with the surrounding magnetic field lines, which are being dragged downward by the infalling cloud and exhibit the characteristic “drip” morphology discussed in \citet{Voit_2026}. This morphology is a direct consequence of magnetic draping: as the cloud moves through the magnetized ICM, field lines are swept up and wrapped around its leading edge, forming a coherent magnetized layer that both shields the cloud from ram pressure stripping and, as discussed above, exerts a tension force opposing the infall.

\section{Discussion}
\label{sect:discussion}

\subsection{Limits of our study}
\label{sect:limits}

Although our Lagrangian tracer particle analysis provides insights into the lookback history of cooling gas in the ICM, we emphasize that these results must not be overinterpreted, and that key questions remain unaddressed. In particular, while our statistical analysis robustly identifies mixing as the dominant thermodynamic pathway for tracers transitioning to the cold phase, the origin of the low-entropy seeds that initiate mixing is not directly constrained by our analysis. In the purely hydrodynamical run, visual inspection suggests that in-situ cooling is responsible for the formation of the low-entropy clumps that subsequently grow through mixing. However, because each clump acquires far more tracers through mixing than were initially present in the cooling cell, the tracer population associated with the initial condensation event is overwhelmed by the mixed component. The importance of in-situ cooling for enabling clump formation is therefore likely underrepresented in our statistics, and caution is warranted when interpreting the relative contributions of the two channels. We note also that tracers are injected once the system has reached a quasi-steady state at $t=2.1$ Gyr, after multiple generations of filaments and clumps have already formed. This choice ensures that the imprint of the initial conditions has largely been erased, so that the statistical signals we extract are not biased by transient features of the initial setup. As a result, we cannot unambiguously trace the ultimate origin of the cold gas reservoir, which initially forms through purely in-situ cooling. This question, tied to the earliest phases of cooling and AGN activity, is better addressed by cosmological simulations \citep[see e.g.,][]{Rohr_2025,Staffehl_2025}; our analysis focuses on the dynamics of cold gas once such a reservoir is established.

A related limitation concerns the absolute mass budget of mixing. Our analysis identifies tracers that are accreted onto pre-existing cold seeds, but does not directly quantify how much mass is supplied through mixing relative to the initial seed mass. Since tracer particles are massless, this quantity is not straightforwardly accessible. The two isolated clouds presented in Sect.~\ref{sect:ballistic} (see also Fig.~\ref{fig:Mass_vs_t}) show no evidence of coagulation or additional in-situ cooling events, and appear to grow solely by collecting ambient tracers along their trajectory, suggesting that mixed gas accounts for more than $95\%$ of their final mass. However, it remains unclear how well this extrapolates to the full clump population, and whether any systematic difference exists between the hydrodynamical and MHD runs in this regard.

Establishing causal relationships between the magnetic field and the kinematic and thermodynamic divergence identified in Sect.~\ref{sect:globaltracers} remains equally challenging. While our analysis robustly establishes that cooling tracers in the MHD run begin to deviate from the background ICM significantly earlier than in the hydrodynamical run, the causal origin of this divergence remains unclear. It is possible that magnetic fields directly promote condensation by altering the local thermodynamic conditions of the gas, but it is equally possible that this signal reflects a selection bias: in the MHD run, cold gas may preferentially collect from regions of the ICM that are already characterised by distinct dynamical conditions (such as the vicinity of infalling filaments) that have no counterpart in the hydrodynamical run. Disentangling these two interpretations would require a more controlled analysis, likely combining targeted idealised simulations with a careful characterisation of the local ICM environment at the time of tracer accretion, and is beyond the scope of the present study. Nevertheless, our results are encouraging: the clear and systematic differences identified between the two runs point to physical processes likely tied to cold gas survival and growth that warrant dedicated follow-up investigation.

Finally, we emphasize that the spatial resolution of our simulations $\Delta x \sim \mathcal{O}(100\,\rm pc$) implies that most cold structures are marginally or entirely unresolved, particularly in the purely hydrodynamical run where the absence of magnetic draping leaves clouds more susceptible to shattering. This is a potentially important limitation, as the mixing efficiency, survival, and growth rate of cold clouds are known to be sensitive to numerical resolution \citep[e.g.,][]{Leary_2026}.

\subsection{Comparison with previous work}

The condensation of hot plasma into neutral and molecular gas has been the subject of substantial work spanning a variety of systems, including supernova remnants, the circumgalactic medium, and the intracluster medium. Here, we restrict ourselves to a brief comparison with recent work focused on the dynamics of the intracluster and circumgalactic media. In particular, our results for the pure hydrodynamic simulation are consistent with those of \citet{Sotira_2026}, who used Eulerian tracer fluids to study the conditions favorable to condensation in a similar Perseus-like cluster. They reported a characteristic timescale of $\sim 30$ Myr and a comparatively prominent role of compression relative to vorticity. Our analysis, based on Lagrangian tracer particles sampling the full simulation volume without any a priori assumptions on the nature of the condensing gas, qualitatively supports these conclusions. As shown in Figs.~\ref{fig:full_vorti_comp}, the deviations in vorticity and compression of the condensing population from the non-condensing background become significant only within the last 30--50 Myr prior to condensation in the pure hydrodynamic case. We emphasize, however, that our MHD run exhibits significant departures from this picture: vorticity coupled with magnetic field amplification appears to play a central role, imprinted over lookback timescales far exceeding those found in the hydrodynamic run. These results strongly advocate for the inclusion of magnetic fields in future simulations of the multiphase intracluster medium, and suggest that further parametric studies are needed to better disentangle the key physical processes at play between the pure hydrodynamic and MHD regimes.

Our results are also in good agreement with more idealized studies of stratified and turbulent box setups such as those of \citet{Ji_2018} and \citet{Wibking_2025,Wibking_2024}, of which they represent a natural extension toward less idealized configurations. While our results support the key role of magnetic fields in enhancing and facilitating condensation, they also suggest a picture in which AGN activity and the motion of cold gas through the ICM provide an amplification mechanism for the magnetic field strength, likely via a small-scale dynamo (see Figs.~\ref{fig:vorticity_vs_mag} and \ref{fig:vorticity_mag_filcloud}).

Finally, our infall velocity analysis for both simulation runs is consistent with more idealized studies assessing the role of magnetic tension on the terminal velocity of infalling cold clouds in the CGM. \citet{Kaul_2025} found that magnetic field configurations orthogonal to the infall direction significantly reduce the terminal velocity of CGM clouds through magnetic draping and the associated tension force, while field lines aligned with the infall direction have a  weaker effect on cloud dynamics \citep[see also e.g.,][]{Dursi_2008}. Both Figs.~\ref{fig:acceleration} and \ref{fig:HVC_mag} are consistent with the orthogonal field scenario, and demonstrate that magnetic draping takes place in our non-idealized simulation, where the ICM magnetic field topology is set self-consistently by the plasma dynamics rather than imposed by hand. Interestingly, spatially isolated cold gas clouds disconnected from the central filamentary network have been observed in several nearby cool-core clusters, including the Perseus cluster, where discrete H$\alpha$- and CO-emitting clouds have been detected at projected distances of several tens of kpc from the brightest cluster galaxy \citep[e.g., the Northern filament system in NGC 1275;][]{Fabian_2003, Fabian_2008}.

\subsection{Prospects for future work}

The results from our MHD simulation leave several questions open for future investigation. First, visual inspection of Fig.~\ref{fig:mainfig} and its associated video suggests that the cold gas cycle differs substantially between the two runs. In the hydro run, cold clumps appear to originate predominantly from in-situ condensation, whereas in MHD the dominant pathway involves the shredding of infalling filaments and potential recycling of cold material through AGN-driven uplift. Quantifying these pathways and comparing the typical lifetime of individual cold gas parcels in hydro and MHD would be a natural extension of this work. Second, the effect of magnetic fields on cloud growth remains to be quantified. While we have shown that the Lagrangian histories of cooling gas parcels differ significantly between the two runs, the complexity of the cluster environment prevents a definitive assessment of whether magnetic fields primarily enhance the net accretion of hot material onto cold clouds, promote their survival \citep{Bruggen_2023,Jennings_2023,Hidalgo_2024,Mohapatra_2025}, or both.

Furthermore, Figs.~\ref{fig:vorticity_vs_mag} and \ref{fig:vorticity_mag_filcloud} suggest that vorticity and magnetic field amplification begin $\sim 150$ Myr before condensation, with a gentle increase of a factor of $2$--$3$ relative to the background medium up to $ \vert t_{\rm lookback} \vert \sim 20$ Myr, followed by a prompt, non-linear growth in the last tens of Myr prior to cooling. Visual inspection of our simulation suggests that the first phase corresponds to a gradual stirring of the ICM driven by the repeated passage of AGN-driven shocks, while the second is likely tied to the runaway cooling of the condensing gas. However, we cannot strictly separate these two regimes, nor quantify the precise role of the first phase. Interestingly, Fig.~\ref{fig:entropyfluctuations} hints that this early phase may promote low-entropy fluctuations, rendering the gas more thermally unstable and more susceptible to cooling. A further complication arises from the fact that condensation in our simulation proceeds in a clustered fashion: massive filaments assemble progressively from smaller condensing clumps that coagulate into larger structures over time. As a consequence, at short lookback times, a given condensing gas element is likely embedded in an environment already populated by neighboring cold structures. This raises the question of how much of the turbulence experienced by the condensing gas is driven by AGN activity, and how much is locally induced by the motion of surrounding cold clumps through the hot ICM, a contribution that is unlikely to be negligible, and which our current analysis cannot cleanly disentangle.

Finally, our results were obtained in the specific context of an idealised Perseus-like cool-core cluster, and it remains an open question how far they can be extrapolated to other environments. Galaxy groups occupy a regime of shallower gravitational potentials, lower ICM temperatures, and potentially higher plasma beta values, all of which could alter the relative importance of magnetic tension, buoyancy, and AGN-driven turbulence in regulating condensation \citep{Prasad_2025}. Similarly, the circumgalactic medium of spiral galaxies, while sharing some phenomenological similarities with cool-core cluster physics, differs substantially in its thermodynamic state, magnetic field strength, and feedback processes. Dedicated simulations across this broader parameter space would help assess the universality of the magnetically regulated condensation picture presented here.

\section{Conclusion}
\label{sect:conclusion}

We presented high-resolution hydrodynamic and MHD simulations of a Perseus-like cool-core cluster in fixed gravity with self-regulated AGN feedback and Lagrangian tracer particles in the GPU-accelerated code \textsc{AthenaPK}. By following the thermodynamic and dynamical histories of gas elements in an unbiased, volume-filling way, we characterized the conditions leading to condensation and assessed the role of magnetic fields throughout the full cycle of cold gas formation and evolution. Our main results are as follows:

\begin{itemize}
    
    \item In both runs, the tracer population transitioning from the hot ($T > 5\times10^6\,\rm K$) to the cold ($T < 10^5\,\rm K$) phase is dominated by hot gas entrained onto cold, low-entropy clumps rather than by in-situ radiative cooling (Figs.~\ref{fig:entropy_fraction} and \ref{fig:closest_neighbor}).
    
    \item The origin of these low-entropy seeds appears to differ qualitatively between the two runs. Visual inspection suggests that in the hydrodynamical run they form mainly via in-situ cooling at the edges of rising cavities, while in the MHD run they originate predominantly from cold gas fragments extracted from infalling filaments by AGN jets and subsequently uplifted into the ICM, with a secondary contribution from in-situ cooling in regions undisturbed by the jets (Fig.~\ref{fig:mainfig}; see also accompanying \href{https://youtu.be/eIKyCpKgpsY}{movie}).

    \item The cooling pathway differs fundamentally between the hydrodynamic and MHD runs. In the purely hydrodynamic case, the condensing tracer population only begins to diverge kinematically from the non-condensing ICM background $\sim 30$ Myr before the cooling transition. In the MHD run, this divergence begins as early as ${\sim}150\,\rm Myr$ before condensation, and is most clearly visible in the evolution of vorticity and magnetic energy (Figs.~\ref{fig:full_vorti_comp} and \ref{fig:vorticity_vs_mag}).

    \item The growth of vorticity and magnetic energy strongly correlate over the full ${\sim}150\,\rm Myr$ pre-condensation window, pointing to a picture in which turbulence driven by AGN activity and / or by the shearing motions around cold gas blobs moving through the ICM contribute to the amplification of magnetic fields prior to condensation (Fig.~\ref{fig:vorticity_vs_mag}).
    
    \item Tracers in the MHD run are characterized by a lower mean turbulent Mach number at the cooling transition, with $\mathcal{M}_{\rm turb,MHD} \approx 0.28$ compared to $\mathcal{M}_{\rm turb,HD} \approx 0.4$ (Fig.~\ref{fig:full_vorti_comp}).
    
    \item Complementing the global tracer analysis, we examine the pre-condensation evolution of tracers associated with individual cold structures identified in the MHD run, including a massive filament forming at the edge of a rising cavity and two isolated ballistic clouds (BCs) condensing at larger cluster-centric radii. While the filament broadly follows the global tracer population, the two BCs exhibit a more abrupt transition, closer to the cooling pathway observed in the purely hydrodynamic run (Fig.~\ref{fig:vorticity_mag_filcloud}).

    \item We then investigate the subsequent evolution of these structures after condensation, focusing on the dynamics of the cold tracers. In this phase, magnetic tension provides a significant deceleration mechanism for the infalling BCs: a decomposition of the radial acceleration terms shows that magnetic tension can counteract up to $\sim 100\%$ of the gravitational acceleration in the final $20$--$30$ kpc of infall, far exceeding the contribution from ram pressure (Figs.~\ref{fig:drag} and \ref{fig:acceleration}).
\end{itemize}

Overall, our study provides novel insight into the thermodynamic history of gas prior to condensation, and reveals statistically significant differences between hydro and MHD. The causal origin of these differences remains open: one possibility is that magnetic fields, potentially amplified by turbulence, alter the thermodynamic conditions of the hot gas, facilitating its subsequent condensation. We leave the clarification of this mechanism, along with a more detailed assessment of cold gas survival in MHD, to future work.

\begin{acknowledgements}
MF thanks Ish Kaul and Fernando Hidalgo-Pineda for insightful discussions, as well as Jacob Shen for providing help with visualisation packages. MF acknowledges the use of Anthropic's Claude and OpenAI's ChatGPT to improve the phrasing of portions of the manuscript text, and to assist with refactoring and optimization of the analysis code. All AI suggestions were reviewed, validated, and thoroughly tested by the author before adoption. \\
MB and MF acknowledge funding by the Deutsche Forschungsgemeinschaft (DFG, German Research Foundation) under Germany's Excellence Strategy -- EXC 2121 ``Quantum Universe'' --  390833306 and project number 443220636 (DFG research unit FOR 5195: "Relativistic Jets in Active Galaxies"). 
PG acknowledges funding by the Deutsche Forschungsgemeinschaft (DFG) – 555983577. This research was supported in part by grant NSF PHY-2309135 to the Kavli Institute for Theoretical Physics (KITP).\\
BWO acknowledges support from NSF grants \#1908109 and \#2106575, NASA ATP grants NNX15AP39G and 80NSSC18K1105, and NASA TCAN grant 80NSSC21K1053. \\
This research used resources of the Oak Ridge Leadership Computing Facility at the Oak Ridge National Laboratory, which is supported by the Office of Science of the U.S. Department of Energy under Contract No. DE-AC05-00OR22725. These resources were provided as part of the DOE INCITE Leadership Computing Program under allocation AST-146 (PI: Brian W. O'Shea).\\
The authors also gratefully acknowledge the Gauss Centre for Supercomputing e.V. (www.gauss-centre.eu) for funding this project by providing computing time through the John von Neumann Institute for Computing (NIC) on the GCS Supercomputer JUWELS at J\"ulich Supercomputing Centre (JSC). \\
All simulations were performed using the public MHD code {\href{https://github.com/parthenon-hpc-lab/athenapk}{\scshape{AthenaPK}}}, which makes use of the \href{https://github.com/kokkos/kokkos}{{\scshape{Kokkos}}} \citep{Kokkos} library and the \href{https://github.com/parthenon-hpc-lab/parthenon}{{\scshape{Parthenon}}} adaptive mesh refinement framework \citep{Parthenon}. All data analysis was performed with \href{https://yt-project.org/}{{\scshape{yt}}} \citep{Turk_2011,yt4}, \href{https://visit-dav.github.io/visit-website/}{{\scshape{VisIt}}} \citep{visit}, \href{https://matplotlib.org/}{{\scshape{Matplotlib}}} \citep{Matplotlib}, \href{https://numpy.org/}{{\scshape{Numpy}}} \citep{Numpy}, \href{https://seaborn.pydata.org/}{{\scshape{Seaborn}}} \citep{Seaborn} and \href{https://cmasher.readthedocs.io/}{{\scshape{CMasher}}} \citep{cmasher}. We thank their authors for making these software and packages publicly available.
\end{acknowledgements}

\bibliographystyle{aa}
\bibliography{references}

\begin{appendix}

\section{Evolution of the BCs properties}

\begin{figure*}[h!]
\includegraphics[width=\textwidth]{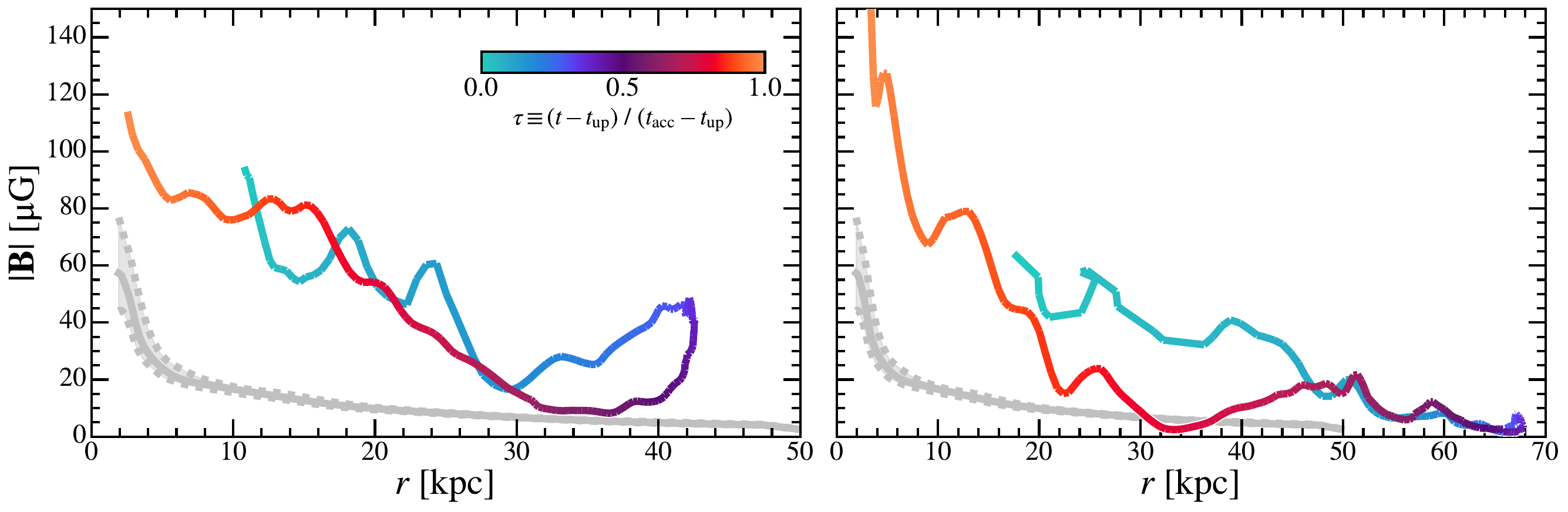}
    \caption{Mass-weighted magnetic field strength in the two infalling ballistics clouds, BC\#1 (left) and BC\#2 (right), shown as a function of their radial distance from the center. The radially-averaged magnetic field strength profile of the hot ($5\times 10^6 \leq T \leq 10^8 \, \rm{K}$) phase is also indicated with the solid grey line and associated time variation.}
    \label{fig:B_norm}
\end{figure*}

Fig.~\ref{fig:B_norm} shows the mass-weighted magnetic field strength in the two ballistic clouds as a function of radial distance from the cluster centre. At the apex of their trajectory, the field strength in both clouds decreases noticeably, consistent with an expansion of the cold gas volume in response to the lower ambient thermal pressure at larger radii. During the infall phase, the field strength increases sharply, which we attribute primarily to the entrainment and stretching of magnetic field lines by the infalling cloud (see also Fig.~\ref{fig:HVC_mag}). In the later stages of infall, the magnetic field strength in the cold phase reaches approximately 4--5 times that of the surrounding hot ICM, reflecting the progressive amplification that accompanies the cloud's acceleration toward the cluster core.

\begin{figure*}[h!]
\includegraphics[width=\textwidth]{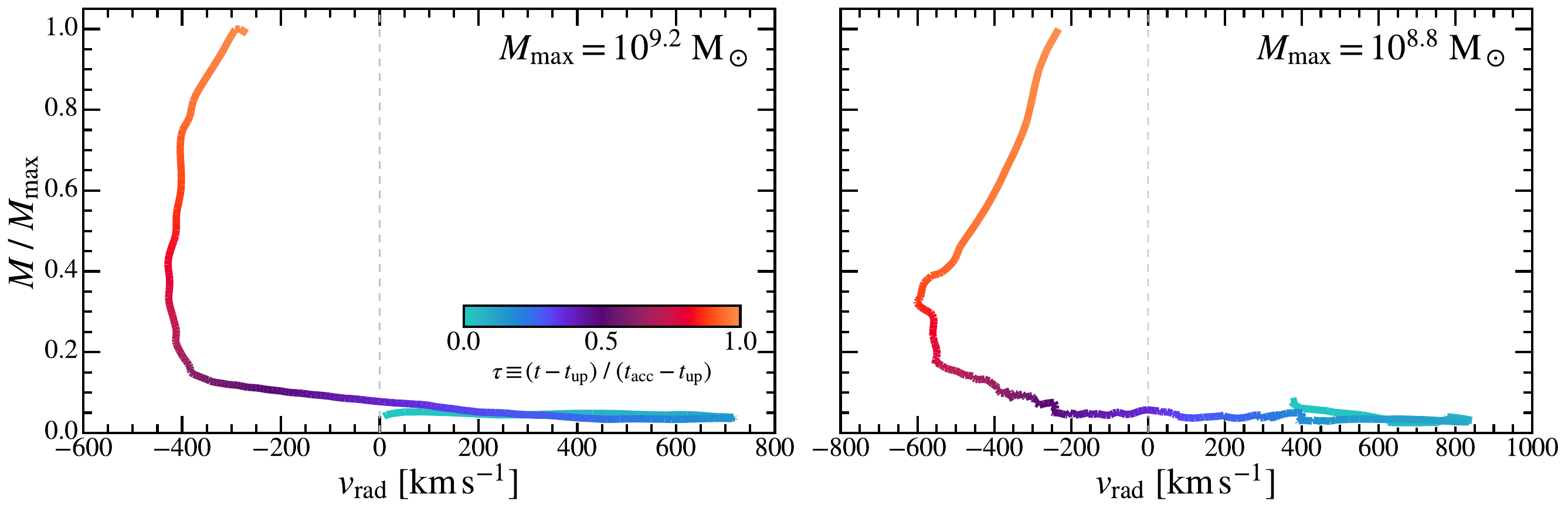}
    \caption{Same as Fig.~\ref{fig:B_norm}, showing the total cold gas mass as a function of mass-weighted radial velocity.}
    \label{fig:Mass_vs_t}
\end{figure*}

Fig.~\ref{fig:Mass_vs_t} reveals an asymmetry in the mass growth history of the two clouds. During the uplift phase and near the apocenter, the total cold gas mass remains approximately constant. A natural explanation is that the clouds spend this phase embedded in the underdense jet cocoon, where the reservoir of hot gas available for entrainment and cooling is limited. The bulk of the mass growth occurs during the infall phase, concentrated around the epoch of peak infall velocity.

\section{Radial profiles of ICM turbulent motions}
\label{app:additionaldiag}

\begin{figure}[h!]
\includegraphics[width=0.5\textwidth]{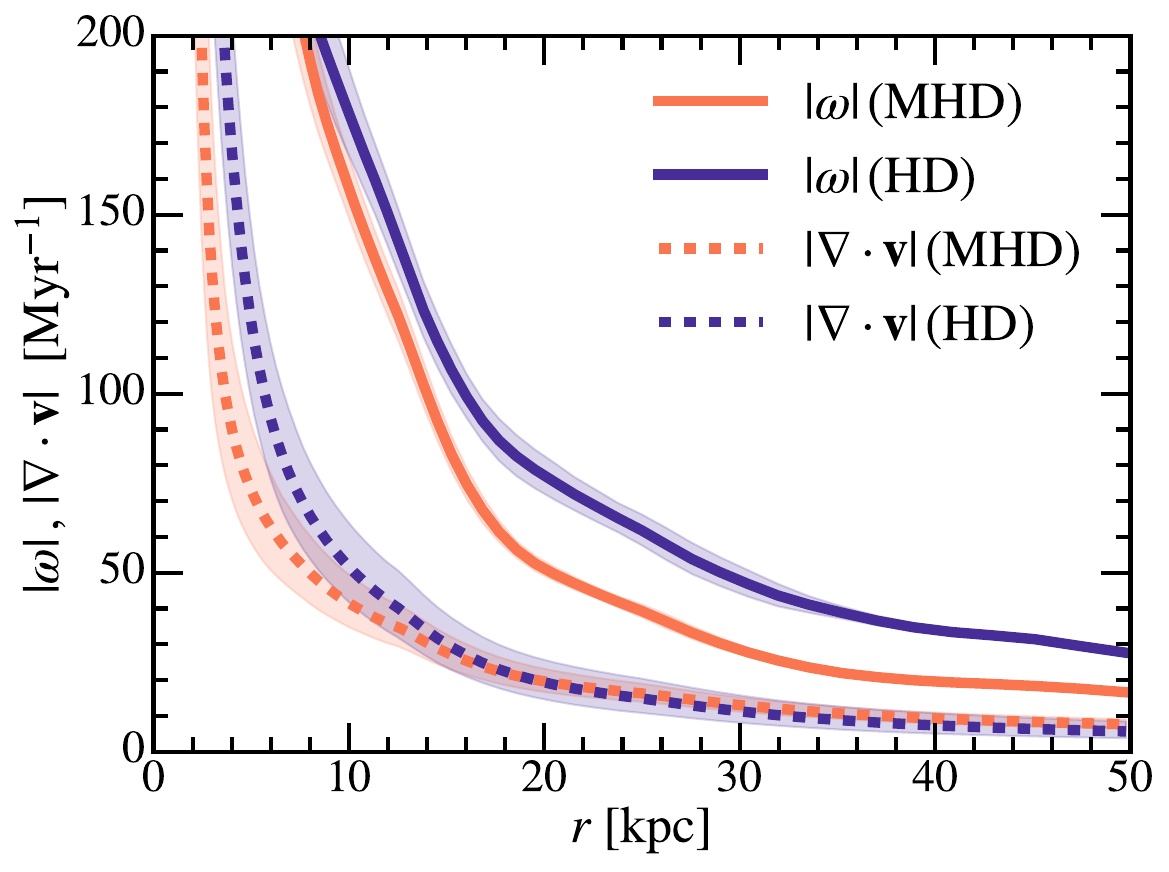}
    \caption{Time-averaged radial profiles of vorticity (solid lines) and compressive (dashed lines) motions in the MHD (orange lines) and hydrodynamical (blue lines) runs. At each snapshot, the quantities are computed from gas cells and then averaged in each radial bin; solid lines show vorticity and dashed lines show compression, with MHD in orange and pure hydro in blue.}
    \label{fig:ICM_profiles}
\end{figure}

\red{Fig.~\ref{fig:ICM_profiles} shows the time-averaged radial profiles of vorticity and compressive motions in the MHD and HD runs, computed from gas cells at each snapshot and averaged within radial bins. These are the same profiles used to normalize the tracer quantities (e.g. the normalized vorticity $\widetilde{\vert \omega \vert}$). For radii larger than $\sim 15$ kpc, the HD run exhibits roughly twice the vorticity of the MHD run, suggesting that magnetic fields suppress rotational motions in the outer ICM — consistent with magnetic tension resisting the bending of field lines associated with vortical flows. The compressive motion profiles are broadly similar between the two runs at all radii.}

\section{Lagrangian tracer particles}
\label{app:tracersmethod}
Due to the Eulerian nature of \athenapk, individual fluid element histories cannot be directly tracked. To overcome this limitation, we implement Lagrangian tracer particles into {\scshape{AthenaPK}}.

Previous studies have highlighted that the choice of advection method is critical for accurately tracing fluid element trajectories. In particular, advection methods based on velocity field interpolation fail to trace gas motion accurately in turbulent box and cosmological simulations, resulting in mismatches between gas and tracer density fields of up to an order of magnitude over time \citep{Genel_2013}. To address this limitation, alternative advection methods based on mass fluxes have been proposed. In \citet{Genel_2013} \citep[see also][]{Cadiou_2019}, tracer positions are updated using a stochastic approach that assigns probabilities for particles to move to neighboring cells based on mass fluxes, which we implement into \athenapk. Hereafter, we summarize the main steps of this method.

Tracers occupy cell centers and can move to neighboring cells at each timestep. For each cell $i$, the displacement probability is $p_{\rm gas} = \Delta M/M_i$, where $\Delta M$ is the total mass endowed by fluxes exiting each cell. For each tracer, a random number $r$ is drawn: if $r < p_{\rm gas}$, the tracer is selected to move. The destination is determined by drawing another random number $r'$ and iterating through neighbors $j$ with positive flux. If $r' < p_j$ (where $p_j = \max(\Delta M_{ij}/\Delta M, 0)$), the tracer moves to cell $j$; otherwise $r'$ is decreased by $r' \leftarrow r' - p_j$ and the next neighbor is tested. When crossing a refinement boundary into a finer region, the tracer is randomly assigned to one of the child cells with probability $p = M_i/M_0$. When crossing into a coarser region, the tracer moves to the parent cell's center \citep[see][for a more detailed description]{Cadiou_2019}. We implemented other advection methods based on velocity field \citep[e.g.][]{Price_2010} and face-centered velocities derived from fluxes interpolation \citep{Tiede_2022}. A comparison between these methods is presented in Appendix~\ref{app:tests}. We store cell-centered grid quantities, in particular, the position, velocity and magnetic vectors, the mass density, the thermal pressure, as well as the acceleration terms and additional quantities summarized in Table~\ref{table:globalquant}.

\section{Tracer advection tests}
\label{app:tests}

\begin{figure}[h!]
\includegraphics[width=0.5\textwidth]{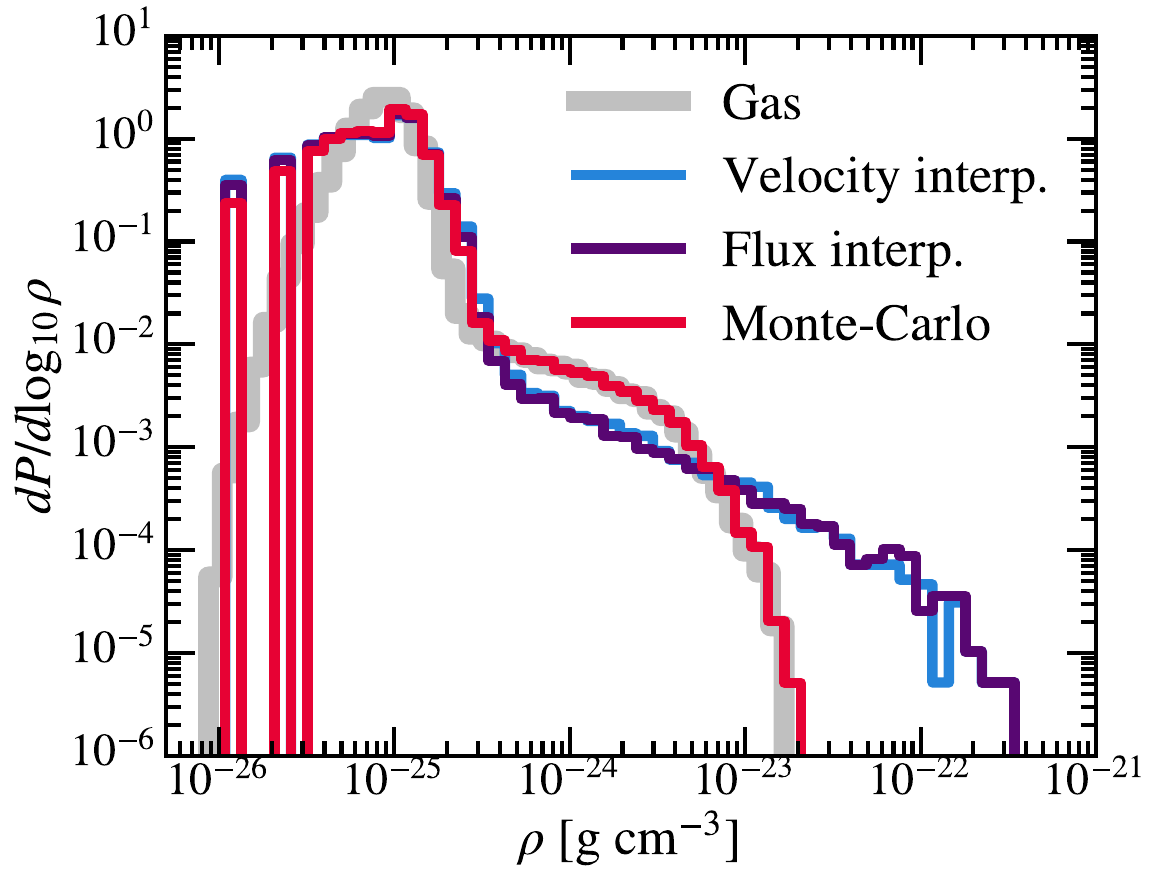}
    \caption{Comparison of gas and tracer density distributions for three different advection schemes. The Monte-Carlo method accurately captures the multiphase structure of the gas, whereas interpolation-based methods show divergences in the high-density tails.}
    \label{fig:appendix_tracer_pdf}
\end{figure}

In order to verify the accuracy of our advection implementation (see Sect.~\ref{sect:method}), we run a turbulent box simulation of width 100 kpc including radiative cooling and tracer particles. We initially seed $N_{\rm{tpc}} = 8$ tracers per cell and evolve the simulation for two turnover times. Fig.~\ref{fig:appendix_tracer_pdf} shows the distribution of tracer mass density as a function of gas density, assuming that each tracer carries a passive mass parameter, $M_{\rm{tr}}$, defined as the total gas mass in the box divided by the total number of tracers, $N_{\rm{cell}} \times N_{\rm{tpc}}$. The three colored lines correspond to three different advection schemes: cell-centered velocity interpolation \citep[e.g.,][]{Price_2010}, face-centered velocity interpolation \citep[e.g.,][]{Tiede_2022}, and a Monte-Carlo approach \citep[e.g.,][]{Genel_2013, Cadiou_2019}. The grey solid line represents the gas density PDF computed directly from the gas cells. As evident, the two velocity-based methods are less accurate in reproducing the density field accurately: they under-sample the $\rho\sim 2\times 10^{-24}$ g cm$^{-3}$ and over-sample the cold phase ($\rho \gtrsim 10^{-23}$ g cm$^{-3}$). In contrast, the Monte-Carlo method closely matches the gas density PDF across the full density range down to tracer mass resolution limit.

\end{appendix}

\end{document}